\documentclass[aps,prc,preprint,tightenlines,groupedaddress,nofootinbib,showpacs,preprintnumbers,amsmath,amssymb,superscriptaddress]{revtex4} 
\usepackage{graphicx}
\usepackage{dcolumn}
\usepackage{bm}
\usepackage{amsmath} 
 
\begin{document} 
 
\def\R{\right} \def\L{\left} \def\Sp{\quad} \def\Sp2{\qquad} 
 
\title{Quasielastic neutrino-nucleus scattering
} 


\author{B. I. S. van der Ventel}
\email{bventel@sun.ac.za}
\affiliation{Department of Physics, University of Stellenbosch,
Stellenbosch 7600, South Africa}

\author{J. Piekarewicz}
\email{jorgep@csit.fsu.edu}
\affiliation{Department of Physics, Florida State University, Tallahassee 
FL 32306, USA}
\date{\today} 
 
\begin{abstract} 
We study the sensitivity of neutral-current neutrino-nucleus
scattering to the strange-quark content of the axial-vector 
form factor of the nucleon. A model-independent formalism for 
this reaction is developed in terms of eight nuclear structure 
functions. Taking advantage of the insensitivity of the ratio 
of proton $(\nu,\nu' p)$ to neutron $(\nu,\nu' n)$ yields to 
distortion effects, we compute all structure functions in a 
relativistic plane wave impulse approximation approach. Further, 
by employing the notion of a bound-state nucleon propagator, 
closed-form, analytic expressions for all nuclear-structure 
functions are developed in terms of an accurately calibrated 
relativistic mean-field model. Using a strange-quark contribution 
to the axial-vector form factor of $g_{A}^{s}\!=\!-0.19$, a 
significant enhancement in the proton-to-neutron yields is 
observed relative to one with $g_{A}^{s}\!=\!0$.

\end{abstract} 
 

\maketitle 
\section{Introduction} 
\label{section_introduction} 

Neutrino physics has established itself at the forefront of current
theoretical and experimental research in astro, nuclear and particle
physics. While neutrino-oscillation experiments, which are presently
evolving from the discovery to the precision phase, will remain at 
the center of most investigations, a variety of other interesting
(non-oscillation) physics topics may be studied in parallel. A prime
example of such a paradigm is the recently commissioned MiniBooNE
experiment at Fermilab. While MiniBooNE's primary goal is to confirm
the neutrino-oscillation experiment at the Liquid Scintillator
Neutrino Detector (LSND) at the Los Alamos National 
Laboratory~\cite{Athanassopoulos_PRL75_95} this unique facility is 
ideal for the study of supernova neutrinos, neutrino-nucleus
scattering, and hadronic structure. In this contribution we focus on 
hadronic structure, in part as a response to the Fermilab Intense 
Neutrino Scattering Experiment (FINeSE) initiative that aims to
measure the strange-quark contribution to the spin of the proton via 
neutral-current elastic scattering~\cite{Potterveld_02}.

A measurement of the spin asymmetry in deep-inelastic scattering of
polarized muons on polarized protons by the European Muon
Collaboration~\cite{Ashman_NPB328_89} revealed a disagreement with the
Ellis-Jaffe sum rule~\cite{Ellis_PRD9_74} in an approach that assumed
that only up and down quarks (and antiquarks) contribute to the
proton spin. This was one of the first indications that hidden flavor
in the nucleon may play an important role in the determination of the
spin structure of the proton.  Experimentally, the spin structure of
the proton is also accessible via parity-violating electron
scattering. Unfortunately, large radiative
corrections~\cite{Musolf_PLB242_90,Musolf_PhysRep239_94} as well as
nuclear-structure effects~\cite{Horowitz_PRC47_93} hinder the
extraction of strange-quark information. A complementary experimental
technique that may be used effectively to study the spin structure of
the proton is elastic neutrino-proton scattering.  The advantages of
neutral-current neutrino-proton scattering over parity-violating
electron scattering and deep-inelastic scattering are well documented
in the literature~\cite{Kaplan_NPB310_88,Potterveld_02}.  Two notable
examples are: (i) the insensitivity of the extraction of the
strange-quark contribution to the use of (broken) SU(3)-flavor
symmetry~\cite{Kaplan_NPB310_88} and (ii) the absence of radiative
corrections in neutral-current neutrino
scattering~\cite{Musolf_PhysRep239_94,Potterveld_02}.  The quark
structure of the nucleon may be investigated in a particularly clean
fashion by evaluating matrix elements of suitable quark-current
operators between single nucleon
states~\cite{Musolf_PhysRep239_94,Anselmino_PhysRep261_95,
Lampe_PhysRep332_2000,Alberico_PhysRep358_2002}. This is 
because quark-current operators may be written in terms of the
fundamental couplings of the quarks to the $Z^{0}$-boson, which are
fully prescribed in the Standard Model.  Further, the weak neutral
current of the nucleon may be parametrized on completely general
ground in terms of two vector and one axial form factors (an
additional induced pseudoscalar form factor is present but its
contribution vanishes in the limit of a zero neutrino mass).  In
particular, the axial-vector form factor may be split into a
non-strange contribution, that may be determined from nuclear beta
decay, and a strange contribution proportional to the fraction of the
nucleon spin carried by the strange quarks~\cite{Garvey_PRC48_93}.
Thus, the axial-vector form factor is crucial to understanding the
role played by strange quarks in determining the properties of the
proton and represents the main focus of this contribution.

A measurement of neutrino-proton and antineutrino-proton elastic
scattering at the Brookhaven National Laboratory (BNL) reported a
non-zero value for the strange form factors of the
nucleon~\cite{Ahrens_PRD35_87}. However, these results must be treated
with caution as the value of the axial mass $M_{A}$ and $g_{A}^{s}$
are strongly correlated~\cite{Garvey_PRC48_93}. Another point of
concern in the BNL experiment was that 80\% of the events involved the
scattering of a neutrino off carbon atoms and only 20\% where from
free protons. Before any firm conclusion may be reached, it is
therefore necessary to understand nuclear-structure effects.
Unfortunately, scattering off a nucleus introduces its own
complications. These include: (i) questions concerning
finite-density effects, such as possible modifications to the nucleon
properties in the nuclear medium and (ii) conventional
nuclear-structure effects, such as binding-energy corrections and
Fermi motion. The first relativistic description of neutrino-nucleus
scattering addressing these complications was presented in
Ref.~\cite{Horowitz_PRC48_93}, where the authors employed a
relativistic Fermi-gas model for the target nucleus. Nuclear-structure
effects were further investigated in Ref.~\cite{Barbaro_PRC54_96} by
considering, in addition to a relativistic Fermi-gas model, a
description of the bound nucleon in terms of harmonic oscillator
wavefunctions. In general it was found that nuclear-structure effects
and final-state interactions~\cite{Garvey_PRC48_93a} can have a
substantial effect on the individual cross sections. However, as
originally suggested in Ref.~\cite{Garvey_PLB289_92}, the {\it ratio} 
of proton to neutron yields is largely insensitive to these 
effects~\cite{Horowitz_PRC48_93,Alberico_NPA623_97}.

In this work we also present a relativistic description of
neutrino-nucleus scattering, but employing bound-state wavefunctions
obtained from an accurately calibrated mean-field
model~\cite{Lalazissis_PRC55_97}.  Further, the impulse approximation
is assumed, that is, the neutrino-nucleon interaction is assumed
unchanged in the nuclear medium. Finally, as the ultimate aim of this
project is to compute {\it ratios} of cross sections, distortion
effects on the ejectile nucleon will be neglected. As we will show
later, this leads to a great simplification in the calculation of all
relevant quantities. The paper has been organized as follows. The
formalism is presented in Sec.~\ref{section_formalism}, followed by
results and conclusions in Sec.~\ref{section_results} and
Sec.~\ref{section_summary}, respectively.

\maketitle
\section{Formalism}
\label{section_formalism}

In this section the formalism for the relativistic description of
(neutral-current) neutrino-nucleus scattering will be presented.  
In particular, it will be shown in Sec.~\ref{section_xsection} that 
the cross section can be written as a contraction between leptonic 
and hadronic tensors. In turn, by relying exclusively on fundamental
principles, the hadronic tensor will be decomposed in terms of a set
of invariant structure functions (see
Sec~\ref{section_hadronic_tensor}). Thus, the formalism is
model-independent. Yet to determine the structure function, and
ultimately the cross section, one must rely on a model. This will be
discussed in Sec.~\ref{section_calc_hadron}.

\subsection{Cross section in terms of leptonic and hadronic tensors}
\label{section_xsection}

\begin{figure}
\includegraphics[height=8cm,angle=0]{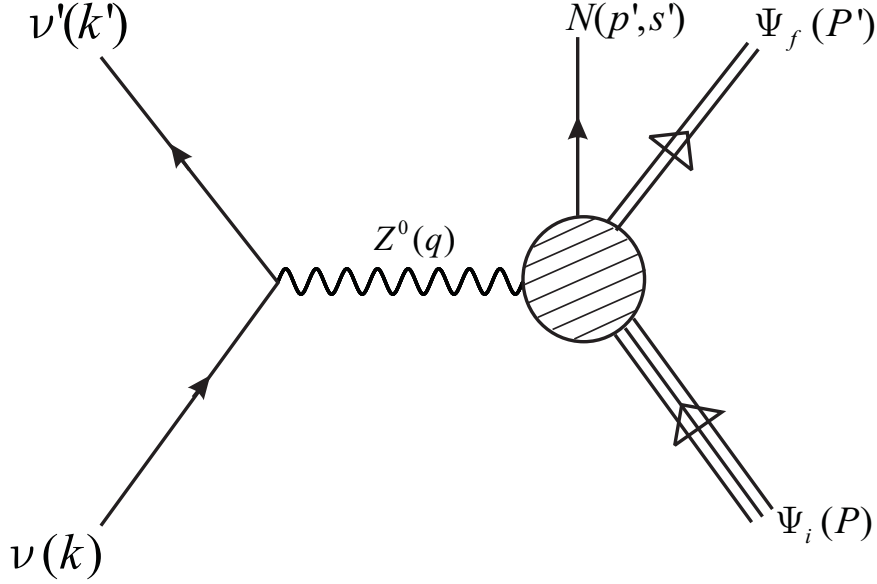}
 \caption{Lowest-order Feynman diagram for the knockout of a
          bound nucleon via neutral-current neutrino-nucleus 
	  scattering \label{fig_1}}
\end{figure}

Due to the short-range nature of the weak interaction, the one-boson
exchange approximation provides an excellent description of
neutrino-nucleus scattering. This results in a cross section that
cleanly separates (or factorizes) into leptonic and hadronic
components. The kinematics of the process is depicted in
Fig.~\ref{fig_1}. Here the initial and final neutrino four-momenta are
denoted by $k$ and $k'$, respectively. Further, the reaction proceeds
via the exchange of a virtual $Z^{0}$-boson with four-momentum
$q$. The target and residual nucleus have four-momenta denoted by $P$
and $P'$, respectively. Finally, the ejectile proton has four-momentum
$p'$ and spin component $s'$.

The differential cross section can now be defined in terms of these 
kinematic variables and the transition matrix element ${\cal M}$ as 
follows:
\begin{eqnarray}
\label{eq_1}
 d \sigma & = &
 \displaystyle \frac{(2 \pi)^{4} \, \delta(k+P-k'-p'-P' \, )}
 {| {\bf v}_{1} - {\bf v}_{2} \, |} \,
 \displaystyle \frac{d^{3}k'}{(2 \pi)^{3}} \;
 \displaystyle \frac{d^{3}p'}{(2 \pi)^{3}} \; 
 \displaystyle \frac{d^{3}P'}{(2 \pi)^{3}} \; |{\cal M}|^{2}\;,
\end{eqnarray}
where ${\bf v}_{1}-{\bf v}_{2}$ denotes the initial relative velocity.
The transition-matrix element contains all the dynamical information 
about the reaction and is given, in the conventions of 
Ref.~\cite{Musolf_PhysRep239_94}, by

\begin{equation}
\label{eq_2}
 -i {\cal M} = 
  \left(\overline{\nu}({\bf k}')
        \left[\frac{igM_{Z}}{4M_{W}}
	(\gamma^{\mu}\!-\!\gamma^{\mu}\gamma^{5})\right]
  \nu({\bf k})\right)iD_{\mu\nu}(q)
  \left(\langle p',\!s';\Psi_{f}(P')\Big| 
  \frac{igM_{Z}}{4M_{W}}\hat{J}^{\nu}(q)\Big|\Psi_{i}(P)
  \rangle\right)\;. 
\end{equation}  

Note that in Eq.~(\ref{eq_2}) the initial and final nuclear states are 
denoted by $\Psi_{i}(P)$ and $\Psi_{f} (P')$, respectively. Further, 
$g$ is the weak coupling constant and $\hat{J}_{\mu}(q)$ is the weak 
nuclear current operator, which contains both vector and axial-vector
components. Finally, as only low momentum transfer ($|q^{2}|\ll
M_{Z}^{2}$) scattering will be considered, the following replacement 
is valid:
\begin{equation}
\label{eq_3}
  D_{\mu\nu}(q)=
  \frac{-g_{\mu\nu}+q_{\mu}q_{\nu}/M_{Z}^{2}}{q^{2}-M_{Z}^{2}}
  \longrightarrow  \frac{g_{\mu\nu}}{M_{Z}^{2}}.
\end{equation}
This allows the transition matrix element to be written as
\begin{equation}
\label{eq_4}
 {\cal M} = \frac{G_{F}}{2\sqrt{2}} 
  \left[\overline{\nu}({\bf k}')
       (\gamma^{\mu}\!-\!\gamma^{\mu}\gamma^{5})
       \nu({\bf k})\right]
  \left[\langle p',\!s';\Psi_{f}(P')\Big| 
  \hat{J}^{\nu}(q)\Big|\Psi_{i}(P)\right] \;.
\end{equation}
Note that $G_{F}$ is the Fermi constant for muon decay which
is given by
\begin{equation}
\label{eq_5}
  G_{F}=\frac{g^{2}}{4\sqrt{2}M_{W}^{2}}\simeq 
        1.166\times 10^{-5}~{\rm GeV}^{-2} \;.
\end{equation}

In Eq.~(\ref{eq_2}) and throughout this work plane-wave Dirac spinors 
are defined as follows:
\begin{equation}
\label{eq_7}
 {\cal U}({\bf k},s) = 
 \sqrt{\frac{E_{k}+M}{2E_{k}}} 
 \left(
  \begin{array}{c}
   1 \\
   \displaystyle{
   \frac{\mbox{\boldmath$\sigma$}\cdot{\bf k}}{E_{k}+M}}
  \end{array}
 \right)\chi_{s}\;,
 \quad \Big(E_{k}\equiv\sqrt{{\bf k}^{2}+M^{2}}\Big) \;.
\end{equation}
Note that the above definition corresponds to the normalization
\begin{equation}
\label{eq_8}
 {\cal U}^{\dagger}({\bf k},s){\cal U}({\bf k},s)=\delta_{ss'}\;.
\end{equation}
This (non-covariant) normalization is motivated by the standard choice 
adopted for bound-state spinors (see Sec.~\ref{section_calc_hadron}
and Ref.~\cite{SW86}) which is given by
\begin{equation}
 \int {\cal U}_{\alpha}^{\dagger}({\bf r})\,
      {\cal U}_{\alpha}({\bf r})\,d{\bf r} = 1\;.
\end{equation}
For (massless) neutrinos in the laboratory frame the initial flux 
factor in Eq.~(\ref{eq_1}) is equal to one. Substitution of the
above relations into Eq.~(\ref{eq_1}) leads to the following
expression for the differential cross section:
\begin{equation}
\label{eq_10}
 d\sigma  = \frac{G_{F}^{2}}{8} 
 (2 \pi)^{4} \delta(k + P - k' - p' - P')  
 \displaystyle \frac{d^{3} {\bf k}'}{(2 \pi)^{3}} \;
 \displaystyle \frac{d^{3} {\bf p}'}{(2 \pi)^{3}} \; 
 \displaystyle \frac{d^{3} {\bf P}'}{(2 \pi)^{3}} \;
 \ell_{\mu\nu} W^{\mu\nu} \;,
\end{equation}
where the leptonic tensor is given by
\begin{equation}
\label{eq_11}
 \ell^{\mu\nu}=
  \Big[\overline{\nu}({\bf k}')
       (\gamma^{\mu}\!-\!\gamma^{\mu}\gamma^{5})
       \nu({\bf k})\Big]
  \Big[\overline{\nu}({\bf k}')
       (\gamma^{\nu}\!-\!\gamma^{\nu}\gamma^{5})
       \nu({\bf k})\Big]^{*}
\end{equation}
while the hadronic tensor by
\begin{equation}
\label{eq_12}
 W^{\mu\nu}=
  \left[\langle p',\!s';\Psi_{f}(P')\Big| 
  \hat{J}^{\mu}(q)\Big|\Psi_{i}(P)\right] 
  \left[\langle p',\!s';\Psi_{f}(P')\Big| 
  \hat{J}^{\nu}(q)\Big|\Psi_{i}(P)\right]^{*} \;.
\end{equation}
The integral over ${\bf P}'$ may be performed using the spatial part
of the Dirac delta function. This fixes the three-momentum of the
residual nucleus to be
\begin{equation}
\label{eq_13}
 {\bf P}'= {\bf k}-{\bf k}'-{\bf p}'+{\bf P} 
 \mathop{\longrightarrow}_{\rm lab}
 {\bf q}-{\bf p}' \;,
\end{equation}
where ${\bf q}\!\equiv\!{\bf k}\!-\!{\bf k}'$ is the three-momentum 
transfer to the nucleus. The differential cross section can now be 
written as
\begin{equation}
\label{eq_13b}
 d\sigma = \frac{G_{F}^{2}}{8(2\pi)^{5}}
 d^{3}{\bf k}' d^{3}{\bf p}' 
 \delta(E_{k}\!+\!M_{A}\!-\!E_{k'}\!-\!E_{p'}\!-\!E_{P'})
 \ell_{\mu\nu} W^{\mu\nu} \;.
\end{equation}

\subsection{Differential cross section in terms of 
            nuclear structure functions}
\label{section_hadronic_tensor}

In Sec.~\ref{section_xsection} it has been shown that the differential
cross for nucleon knockout in neutrino-nucleus scattering involves the
contraction between the leptonic tensor $\ell_{\mu\nu}$
[Eq.~(\ref{eq_11})] and the hadronic tensor $W^{\mu\nu}$
[Eq.~(\ref{eq_12})]. In this section both of these quantities will be 
calculated in a model-independent way by introducing a suitable set of
nuclear structure functions.

Starting from Eq.~(\ref{eq_11}) it follows that the leptonic tensor 
may be written as 
\begin{eqnarray}
\label{eq_22}
 \ell^{\mu\nu}&=&{\rm Tr}
  \left[
   (\gamma^{\mu}\!-\!\gamma^{\mu}\gamma^{5})
   \Big(\nu({\bf k})\overline{\nu}({\bf k})\Big)
   (\gamma^{\nu}\!-\!\gamma^{\nu}\gamma^{5})
   \Big(\nu({\bf k}')\overline{\nu}({\bf k}')\Big)
  \right] \nonumber \\ 
 &=&\frac{2}{kk'}
 \Big[
  k^{\mu}k^{\prime\nu}+k^{\prime\mu}k^{\nu}-g^{\mu\nu}k\cdot k'+
  ih \epsilon^{\mu\nu\alpha\beta}k_{\alpha}k'_{\beta}
 \Big]\;,
\end{eqnarray}
where we made use of Eq.~(\ref{eq_7}) to write
\begin{equation}
\label{eq_23}
 \nu({\bf k})\overline{\nu}({\bf k})=
 \frac{\rlap/k}{2k}
 \left[\frac{1}{2}(1-h\gamma^{5})\right] \;,
 \quad (k\equiv{|{\bf k}|})\;.
\end{equation}
Note that in the above expressions $h\!=\!-1$ and $h\!=\!+1$ refer 
to left-handed neutrinos and right-handed antineutrinos, 
respectively. For later convenience, the leptonic tensor can now be 
separated into a symmetric and an antisymmetric part. That is,
\begin{equation}
\label{eq_25}
 \ell^{\mu\nu} \equiv \ell_{S}^{\mu\nu} + 
                      \ell_{A}^{\mu\nu} \;,
\end{equation}
where
\begin{subequations}
\begin{eqnarray}
 \ell_{S}^{\mu\nu} &=& \frac{2}{kk'}
 (k^{\mu}k^{\prime\nu}+k^{\prime\mu}k^{\nu}-
  g^{\mu\nu}k\cdot k') \;, \\
 \ell_{A}^{\mu\nu} &=& \frac{2}{kk'}
 ih \epsilon^{\mu\nu\alpha\beta}k_{\alpha}k'_{\beta} \;.
\end{eqnarray}
\label{eq_26}
\end{subequations}
Note that all remnants of the neutrino helicity resides in the 
term containing the antisymmetric Levi-Civita tensor. It is only 
this antisymmetric component of the leptonic tensor that is sensitive 
to the difference between an incident neutrino or antineutrino beam. 
It then follows from Eq.~(\ref{eq_26}), as the weak neutral currents 
is conserved for massless neutrinos, that
\begin{equation}
\label{eq_27a}
 q_{\mu}\ell^{\mu\nu}=\ell^{\mu\nu}q_{\nu}=0\;,
\end{equation}
where $q^{\mu}\!\equiv\!(\omega,{\bf q})\!=\!
(k^{\mu}\!-\!k^{\prime\mu})$ is the four-momentum transfer to the 
nucleus.

The hadronic tensor is an extremely complicated object as in principle
exact many-body wave functions and operators must be used. Yet it
follows from Eq.~(\ref{eq_12}) that for unpolarized nucleon emission
the hadronic tensor is only a function of three independent
four-momenta: $q^{\mu}$, $P^{\mu}$ and $p^{\prime\mu}$, as the
four-momentum of the recoiling nucleus $P^{\prime\mu}$ is fixed by
four-momentum conservation. The hadronic tensor can therefore be
parametrized in terms of a basis constructed from the following five
tensors: $\left\{q^{\mu}, P^{\mu}, p'^{\prime\mu}, g^{\mu \nu},
\epsilon^{\mu \nu \alpha \beta} \right\}$. This is analogous to the
case of electron scattering but now we are no longer allowed to
invoke either parity invariance or current conservation, as the weak 
interaction violates parity and the axial-vector current is not 
conserved. We start by separating the hadronic tensor into symmetric 
and antisymmetric components. That is,
\begin{equation}
\label{eq_28}
 W^{\mu\nu} \equiv W_{S}^{\mu\nu} + W_{A}^{\mu\nu} \;.
\end{equation}
Using the above-mentioned basis the individual components may be
written as follows:
\begin{subequations}
\begin{eqnarray}
 W_{S}^{\mu\nu} & = &
 W_{1} g^{\mu \nu}+ W_{2} q^{\mu}q^{\nu}+W_{3} P^{\mu}P^{\nu}+
 W_{4}\, p^{\prime\mu}p^{\prime\nu} \nonumber \\
                & + &
 W_{5} (q^{\mu}P^{\nu}\!+\!P^{\mu}q^{\nu})+
 W_{6} (q^{\mu}p^{\prime\nu}\!+p^{\prime\mu}q^{\nu})+
 W_{7} (P^{\mu}p^{\prime\nu}\!+p^{\prime\mu}P^{\nu})\;,
 \label{eq_29} \\
 W_{A}^{\mu\nu} & = &
 W_{8}  (q^{\mu}P^{\nu}\!-\!P^{\mu}q^{\nu}) +
 W_{9}  (q^{\mu}p^{\prime\nu}\!-p^{\prime\mu}q^{\nu})+
 W_{10} (P^{\mu}p^{\prime\nu}\!-p^{\prime\mu}P^{\nu}) \nonumber \\
                & + &
 W_{11} \epsilon^{\mu\nu\alpha\beta}q_{\alpha}P_{\beta} +
 W_{12} \epsilon^{\mu\nu\alpha\beta}q_{\alpha}p'_{\beta} +
 W_{13} \epsilon^{\mu\nu\alpha\beta}P_{\alpha}p^{\prime}_{\beta}\;.
\label{eq_30}
\end{eqnarray}
\end{subequations}
Note that all structure functions are functions of the four
Lorentz-invariant quantities, $q^{\mu}q_{\mu}\!\equiv\!-Q^{2}$,
$q\cdot P$, $q\cdot p^{\prime}$, and $P\cdot p^{\prime}$. The hadronic
tensor for the $(\nu,\nu' p)$ [or $(\nu,\nu' n)$] reaction contains
thirteen independent structure functions. Contrast this, for example,
to the hadronic tensor for $(e,e' p)$ which is fully written in terms
of only five structure functions~\cite{Picklesimer_PRC32_85}. 

We now proceed to evaluate the contraction of the leptonic tensor
with the hadronic tensor. First, we introduce the following 
definition:
\begin{equation}
\label{eq_31}
{\cal F}(k,k';P,p') \equiv \left(\frac{4}{kk'}\right)^{-1}
                           \ell_{\mu\nu} W^{\mu\nu} 
                    ={\cal F}_{S}(k,k';P,p')
                    +{\cal F}_{A}(k,k';P,p')\;,
\end{equation}
where ${\cal F}_{S}$ and ${\cal F}_{A}$ are defined in terms of the
symmetric and antisymmetric components of the leptonic and hadronic 
tensors, respectively. That is,
\begin{subequations}                  
\begin{eqnarray}
\label{eq_32a}
   {\cal F}_{S}(k,k';P,p') &\equiv& 
   \left(\frac{4}{kk'}\right)^{-1}
   \ell_{\mu\nu}^{S} W^{\mu\nu}_{S}
        \nonumber \\  &=& 
   \Bigg(
      -W_{1}(k\cdot k^{\prime})+
       W_{3}\left[(k\cdot P)(k^{\prime}\cdot P)+
            \frac{M_{A}}{2}^{2}(k\cdot k^{\prime})\right]
        \nonumber \\  &+& 
       W_{4}\left[(k\cdot p^{\prime})(k^{\prime}\cdot p^{\prime})+
            \frac{M_{N}^{2}}{2}(k\cdot k^{\prime})\right]
        \nonumber \\  &+&  
       W_{7}\Big[(k\cdot P)(k^{\prime}\cdot p^{\prime})+
	          (k\cdot p^{\prime})(k^{\prime}\cdot P)-
	          (k\cdot k^{\prime})(P\cdot p^{\prime})
	    \Big]
   \Bigg)\;, \\
\label{eq_32b}
   {\cal F}_{A}(k,k';P,p') &\equiv& 
   \left(\frac{4}{kk'}\right)^{-1}
   \ell_{\mu\nu}^{A} W^{\mu\nu}_{A} 
        \nonumber \\  &=& ih
   \Bigg(
       W_{10}\epsilon^{\mu\nu\alpha\beta}
         k_{\mu}k^{\prime}_{\nu}P_{\alpha}p^{\prime}_{\beta}+
       W_{11}(k\cdot k^{\prime})
             \Big(k\cdot P+k^{\prime}\cdot P\Big) 
             \nonumber \\  &+& 
       W_{12}(k\cdot k^{\prime})
             \Big(k\cdot p^{\prime}+k^{\prime}\cdot p^{\prime}\Big)
             \nonumber \\  &+&  
       W_{13}\Big[(k\cdot p^{\prime})(k^{\prime}\cdot P)-
                   (k\cdot P)(k^{\prime}\cdot p^{\prime})
	     \Big]
   \Bigg)\;.
\end{eqnarray}
\end{subequations}
Note that as a result of current conservation, only eight of the
original thirteen structure functions survived the contraction [see
Eq.~(\ref{eq_27a})]. Further, an interesting difference between
neutrino and electron scattering can be seen from
Eq. (\ref{eq_32b}). While for electron scattering one must prepare a
polarized beam to sample the antisymmetric part of the hadronic
tensor~\cite{Picklesimer_PRC32_85}, for neutrinos the polarization
happens by default. Eqs.~(\ref{eq_13b}), (\ref{eq_32a}), and
(\ref{eq_32b}) comprise the principal results for this section. It is
the most general structure possible for nucleon knockout in
neutral-current neutrino-nucleus scattering. It shows that the
differential cross section is completely determined by a set of eight
structure functions multiplied by kinematical factors.

\subsection{Model-dependent calculation of the cross section}
\label{section_calc_hadron}

In the previous section it was shown that the differential cross
section is completely determined by a set of eight structure 
functions. These structure functions parametrize our ignorance about 
strong-interaction physics. In principle, these structure functions 
could be measured through a ``super'' Rosenbluth separation. In 
practice, however, this is beyond realistic expectations. Thus, 
it is not possible to proceed further without an explicit model 
of the hadronic vertex.

First, we focus on some ``kinematical'' approximations that are made 
in order to simplify the argument of the energy conserving delta 
function in Eq.~(\ref{eq_13b}). In the laboratory frame the total 
energy of the residual nucleus is given by
\begin{equation}
\label{eq_14}
 E_{P'} =  \sqrt{{\bf P}^{2}+M_{A-1}^{2}} = 
 \sqrt{({\bf q}-{\bf p}')^{2}+ M_{A-1}^{2}}
 \approx M_{A-1}\;,
\end{equation}
where the last approximation follows in the limit of no recoil
corrections. Further, it is assumed that the energy transfer to 
the nucleus satisfies
\begin{equation}
\label{eq_15}
 \omega=k-k'=E_{p'}-E_{B} \;,
\end{equation}
where $E_{B}$ is the energy of the struck nucleon. This implies
that
\begin{equation}
\label{eq_16}
 E_{P'} =  M_{A} - E_{B} \;,
\end{equation}
and justifies the following replacement in Eq.~(\ref{eq_13b}): 
\begin{equation}
\label{eq_17}
 \delta(E_{k}\!+\!M_{A}\!-\!E_{k'}\!-\!E_{p'}\!-\!E_{P'})
 \longrightarrow 
 \delta(k\!-\!k'\!-\!E_{p'}\!+\!E_{B}) \;.
\end{equation}
Note that we have defined $E_{k}\!=\!|{\bf k}|\!\equiv\!k$ and
$E_{k'}\!=\!|{\bf k}'|\!\equiv\!k'$. 

\begin{figure}[h]
\includegraphics[height=8cm,angle=0]{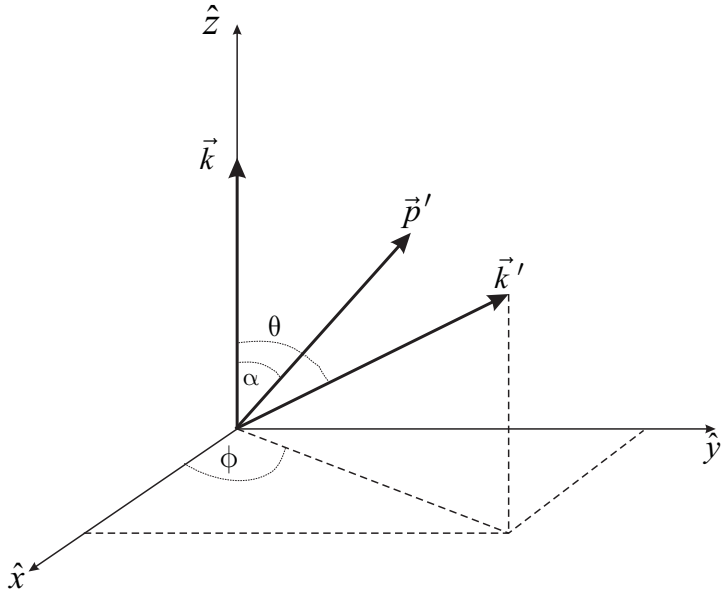}
 \caption{Coordinate axes used to define the angles $\alpha$,
 $\theta$ and $\phi$. \label{fig_2} 
 }
\end{figure}
Consider now the coordinate system shown in Fig.~\ref{fig_2} where
the incoming neutrino defines the $\hat{\bf z}$-axis. The outgoing
nucleon is detected at a scattering angle $\alpha$ relative to the
$\hat{\bf z}$-axis, while the outgoing neutrino, assumed undetected,
has polar and azimuthal angles $\theta$ and $\phi$, respectively.  
The following relations are therefore valid:
\begin{equation}
\label{eq_18}
 d^{3}{\bf k}'= k^{\prime 2}\sin\theta 
                 dk^{\prime}d\theta d\phi 
                 \quad{\rm and}\quad
 d^{3}{\bf p}'=2\pi p'E_{p'}dE_{p'}d(\cos\alpha) \;.
\end{equation}
As the outgoing neutrino will remain undetected, one must integrate 
over $k'$, $\theta$ and $\phi$. Using Eqs.~(\ref{eq_17}) and 
(\ref{eq_18}) the following expression for the differential cross 
section [Eq.~(\ref{eq_13b})] is obtained:
\begin{equation}
 \label{eq_20}
  \frac{d^{2}\sigma}{dE_{p'}d(\cos\alpha)} =
  \left(\frac{G_{F}^{2}}{32\pi^{4}}\right)
  \left(\frac{|{\bf k}^{\prime}||{\bf p}^{\prime}|E_{p'}}
             {|{\bf k}|}\right)
  \int_{0}^{\pi} \sin\theta d\theta \int_{0}^{2\pi}d\phi\,
  {\cal F}(k,k';P,p') \;,
\end{equation}
where the energy conserving delta function constrains the energy of
the outgoing neutrino to 
$|{\bf k}^{\prime}|\!=\!|{\bf k}|\!+\!E_{B}\!-\!E_{p'}$.

\begin{figure}
\includegraphics[height=7cm,angle=0]{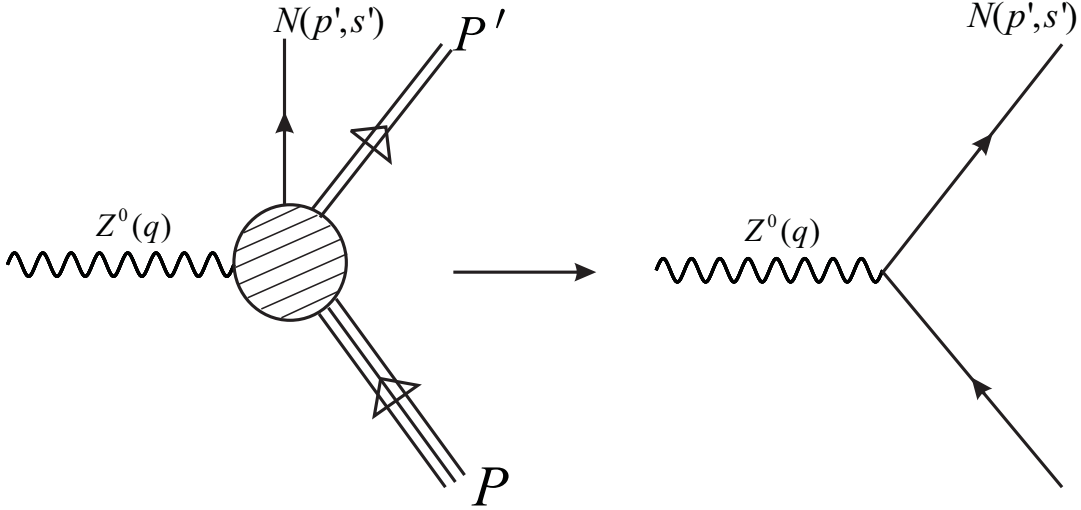}
 \caption{Graphical representation of the approximation employed at the
 hadronic vertex to obtain a tractable form for the matrix element given in
 Eq. (\ref{eq_34}). \label{fig_3}}
\end{figure}

What remains now is to provide an explicit form for the eight
independent structure functions introduced in the previous
section. It follows from Eq.~(\ref{eq_4}) that the dynamical 
information on the hadronic vertex is contained in the following 
matrix element (and its complex conjugate):
\begin{equation}
 \label{eq_34}
  J^{\mu} = \langle p',\!s';\Psi_{f}(P')\Big|
            \hat{J}^{\mu}(q)\Big|\Psi_{i}(P)\rangle \;.
\end{equation}
To obtain a tractable form for this extremely complicated object 
we rely on the approximations depicted in Fig.~\ref{fig_3}, that 
we now address in detail.
First, it is assumed that the $Z^{0}$-boson couples to a single 
bound nucleon. This neglects two- and many-body components of the 
current operator. Second, it is assumed that the detected nucleon 
is the one to which the virtual boson couples to. This neglects 
two- and many-body rescattering processes. Finally, we neglect 
final-state interactions (distortions) of the ejected nucleon. 
While many similar treatments incorporate distortion effects on 
the ejectile (see for example Ref.~\cite{Picklesimer_PRC32_85} 
in the case of electron scattering) concentrating on ratios of 
cross sections makes the formalism largely insensitive to distortion 
effects~\cite{Garvey_PRC48_93}. This will render the hadronic tensor 
analytic. Incorporating the above simplifications yields a 
hadronic matrix element of the following simple form: 
\begin{equation}
\label{eq_37}
 J^{\mu} = \overline{\cal U}({\bf p}',s')
    \hat{J}^{\mu}(q){\cal U}_{\alpha}({\bf p}_{m})\;,
\end{equation}
where $\alpha$ represents a collection of quantum numbers and 
${\bf p}_{m}\!\equiv\!{\bf p}'\!-\!{\bf q}$ is the missing momentum.
Note that the bound-state wavefunction is given by
\begin{equation}
\label{eq_39}
 {\cal U}_{\alpha}({\bf p}) \equiv {\cal U}_{E\kappa m}({\bf p}) =
 \frac{4\pi}{p}(-i)^{l} 
 \left( \begin{array}{c}
         g_{E\kappa}(p) \\
         f_{E\kappa}(p)(\mbox{\boldmath$\sigma$}\cdot\hat{\bf p})
 \end{array}\right) 
 {\cal Y}_{\kappa m}(\hat{\bf p})\;.
\end{equation}
Here $E$ is the bound-state energy, $\kappa$ the (generalized) angular 
momentum, and $m$ the spin projection. Further, $g_{E\kappa}(p)$ and 
$f_{E\kappa}(p)$ are Fourier transforms of the upper and lower
components of the bound-state wavefunction, 
respectively~\cite{Gardner_PRC50_94}.

The impulse approximation is now invoked by assuming that the weak
neutral current for a nucleon in the nuclear medium retains its 
free-space form. That is,
\begin{equation}
\label{eq_40}
 \hat{J}_{\mu}\equiv\hat{J}_{\mu}^{\rm NC} - \hat{J}_{\mu 5}^{\rm NC}
   =\widetilde{F}_{1}(Q^{2})\gamma_{\mu}+ 
   i\widetilde{F}_{2}(Q^{2})\sigma_{\mu\nu}\frac{q^{\nu}}{2M}-
    \widetilde{G}_{A}(Q^{2})\gamma_{\mu}\gamma_{5} \;.
\end{equation}
Here $\widetilde{F}_{1}$ and $\widetilde{F}_{2}$ are Dirac and 
Pauli vector form factors, respectively and $\widetilde{G}_{A}$ 
is the axial form factor. The additional induced pseudoscalar 
form factor (proportional to $q_{\mu}$) does not contribute to 
(massless) neutrino scattering and will be neglected henceforth.
The two vector form factors may be decomposed in terms of the
usual electromagnetic Dirac and Pauli form factors, plus a yet 
undetermined strange-quark contribution. Similarly, the axial vector
form factor consists of a purely isovector contribution, that may be
determined from Gamow-Teller $\beta$-decay rates, and a purely
isoscalar strange-quark contribution. It is the aim of this
contribution to explore the sensitivity of neutral-current
neutrino-nucleus scattering to the strange-quark contribution to the
axial form factor. A detailed discussion of the weak neutral current
[Eq.~(\ref{eq_40})] is given in Appendix~\ref{section_appendix1}.

It is now possible, using Eq.~(\ref{eq_37}), to explicitly calculate
the hadronic tensor defined in Eq. (\ref{eq_12}). Assuming that the 
spin of the outgoing nucleon is not detected, the hadronic tensor may 
be written as
\begin{eqnarray}
 \nonumber
 W^{\mu\nu} & = &\sum_{s'}\sum_{m} 
   \left[\overline{{\cal U }}({\bf p}^{\prime},s^{\prime})
   \hat{J}^{\mu}(q) {\cal U}_{\alpha,m}({\bf p}_{m})\right] 
   \left[\overline{{\cal U }}({\bf p}^{\prime},s^{\prime}) 
   \hat{J}^{\nu}(q) {\cal U}_{\alpha,m}({\bf p}_{m})\right]^{*} \\
\label{eq_54}
 & = &
 {\rm Tr}\left[\hat{J}^{\mu}(q)S_{\alpha}({\bf p}_{m})
         \hat{\bar{J}}^{\nu}(q)S({\bf p}^{\prime})\right]\;,
\end{eqnarray}
where
$\hat{\bar{J}}^{\nu}\!=\!\gamma^{0}(\hat{J}^{\nu})^{\dagger}\gamma^{0}$.
The normalization used in Eq.~(\ref{eq_8}) implies that the standard 
(on-shell) Feynman propagator is given by
\begin{equation}
\label{eq_55}
 S({\bf p}^{\prime})=\sum_{s^{\prime}}
 {\cal U } ({\bf p}^{\prime},s^{\prime}) \,
  \overline{{\cal U }}({\bf p}^{\prime},s^{\prime}) =
  \frac{\rlap/p^{\prime}+M}{2E_{{\bf p}^{\prime}}} \;. 
\end{equation}
Moreover, it has been shown in Ref.~\cite{Gardner_PRC50_94} that the 
following simple identity is valid even in the case of a bound-state 
spinor:
\begin{equation}
\label{eq_56}
  S_{\alpha}({\bf p})\equiv\sum_{m}
               {\cal U}_{\alpha,m}({\bf p}) \,
      \overline{\cal U}_{\alpha,m}({\bf p}) 
   =  (2j+1)({\rlap/{p}}_{\alpha} + M_{\alpha}) \;.
\end{equation}
Note that in the above equation for the ``bound-state'' propagator
mass-, energy-, and momentum-like quantities have been introduced. 
These are given by
\begin{subequations}
\label{eq_57}
\begin{eqnarray}
  M_{\alpha} &=& \left(\frac{\pi}{p^{2}}\right)
                  \Big[g_{\alpha}^{2}(p) -
                       f_{\alpha}^{2}(p)\Big] \;, \\
  E_{\alpha} &=& \left(\frac{\pi}{p^{2}}\right)
                  \Big[g_{\alpha}^{2}(p) +
                       f_{\alpha}^{2}(p)\Big] \;,
 \label{epm} \\
  {\bf p}_{\alpha} &=& \left(\frac{\pi}{p^{2}}\right)
                   \Big[2 g_{\alpha}(p)
                          f_{\alpha}(p)\hat{\bf p}
                   \Big]  \;,
\end{eqnarray}
\end{subequations}
and satisfy the ``on-shell relation''
\begin{equation}
\label{eq_58}
  p_{\alpha}^{2}=E_{\alpha}^{2}-{\bf p}_{\alpha}^{2}
                =M_{\alpha}^{2} \;.
\end{equation}
It now follows from Eqs.~(\ref{eq_55}) and~(\ref{eq_56}) that
the hadronic tensor may be written in the following simple form:
\begin{equation}
\label{eq_61}
 W^{\mu\nu} = \left(\frac{2j+1}{2E_{{\bf p}^{\prime}}}\right)
 {\rm Tr} 
 \left[ 
  \hat{J}^{\mu}(q)({\rlap/{p}}_{\alpha}+M_{\alpha})
  \hat{\bar{J}}^{\nu}(q)(\rlap/p^{\prime}+M) 
 \right]\;.
\end{equation}
The fact that the hadronic tensor may be expressed as a trace over 
Dirac matrices, despite the presence of the bound-state wave function,
greatly simplifies the calculation. It is important to note, however,
that this enormous simplification would have been lost if distortion
effects would have been incorporated in the propagation of the emitted
nucleon. The emphasis on computing ratios of cross sections is the
main justification behind this simplification. We now obtain
\begin{eqnarray}
 \nonumber
 W^{\mu\nu}
 & = & \left(\frac{2j+1}{2E_{{\bf p}^{\prime}}}\right)
 \left[ 
    \widetilde{W}_{1} g^{\mu \nu} + 
    \widetilde{W}_{2} \left(p'^{\mu}p_{\alpha}^{\nu} + 
                       p_{\alpha}^{\mu} p'^{\nu}\right) + 
    \widetilde{W}_{3} \left(q^{\mu}p_{\alpha}^{\nu} + 
                      p_{\alpha}^{\mu}q^{\nu} \right) +
    \widetilde{W}_{4} \left( q^{\mu} p'^{\nu} + 
                      p'^{\mu} q^{\nu} \right) +  \right.\\
\label{eq_62}
 & & \left. \hspace{2.00cm}
  \widetilde{W}_{5} q^{\mu} q^{\nu} + 
  \widetilde{W}_{6} \epsilon^{\mu\nu\sigma\lambda}
                    p_{\alpha,\sigma}p_{\lambda}' +
  \widetilde{W}_{7} \epsilon^{\mu\nu\sigma\lambda} 
                    p_{\alpha,\sigma}q_{\lambda} +
  \widetilde{W}_{8} \epsilon^{\mu\nu\sigma\lambda} 
                    p_{\sigma}'q_{\lambda}
 \right]\;.
\end{eqnarray}
All components of the hadronic tensor are given in terms
of relatively simple expressions that have been collected
in Appendix~\ref{section_appendix2}. These model-dependent
structure functions may be related to the model-independent
ones [see Eqs.~(\ref{eq_29}) and~(\ref{eq_30})] as has been 
shown in Appendix~\ref{section_appendix2}. This concludes the 
formalism for neutral-current neutrino-nucleus scattering.
The explicit forms of the structure functions given in the
appendix can now be used to evaluate the differential cross 
section defined in Eq.~(\ref{eq_20}). 

\section{Results}
\label{section_results}

\begin{figure}
\includegraphics[height=15cm,angle=-90]{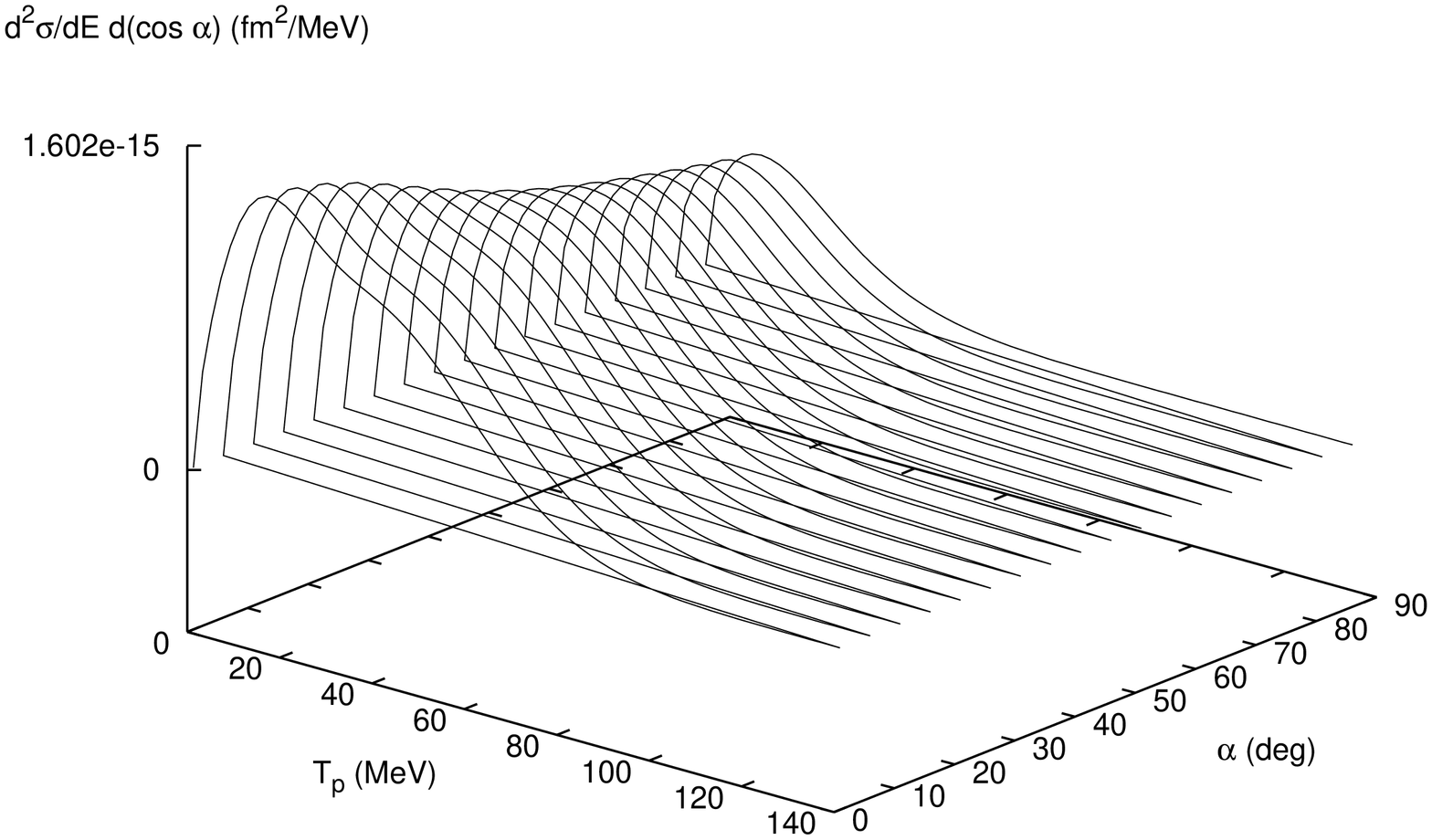}
\caption{Double differential cross section 
 $d^{2} \sigma/dE d (\cos \alpha)$ as a function of the outgoing 
 proton laboratory kinetic energy $T_{p}$ and laboratory scattering 
 angle $\alpha$. The calculation shown is for proton knockout from 
 the $1p^{3/2}$ orbital of $^{12}$C at an incident neutrino energy 
 of 150 MeV.}
 \label{fig_4} 
\end{figure}

\begin{figure}
\includegraphics[height=15cm,angle=-90]{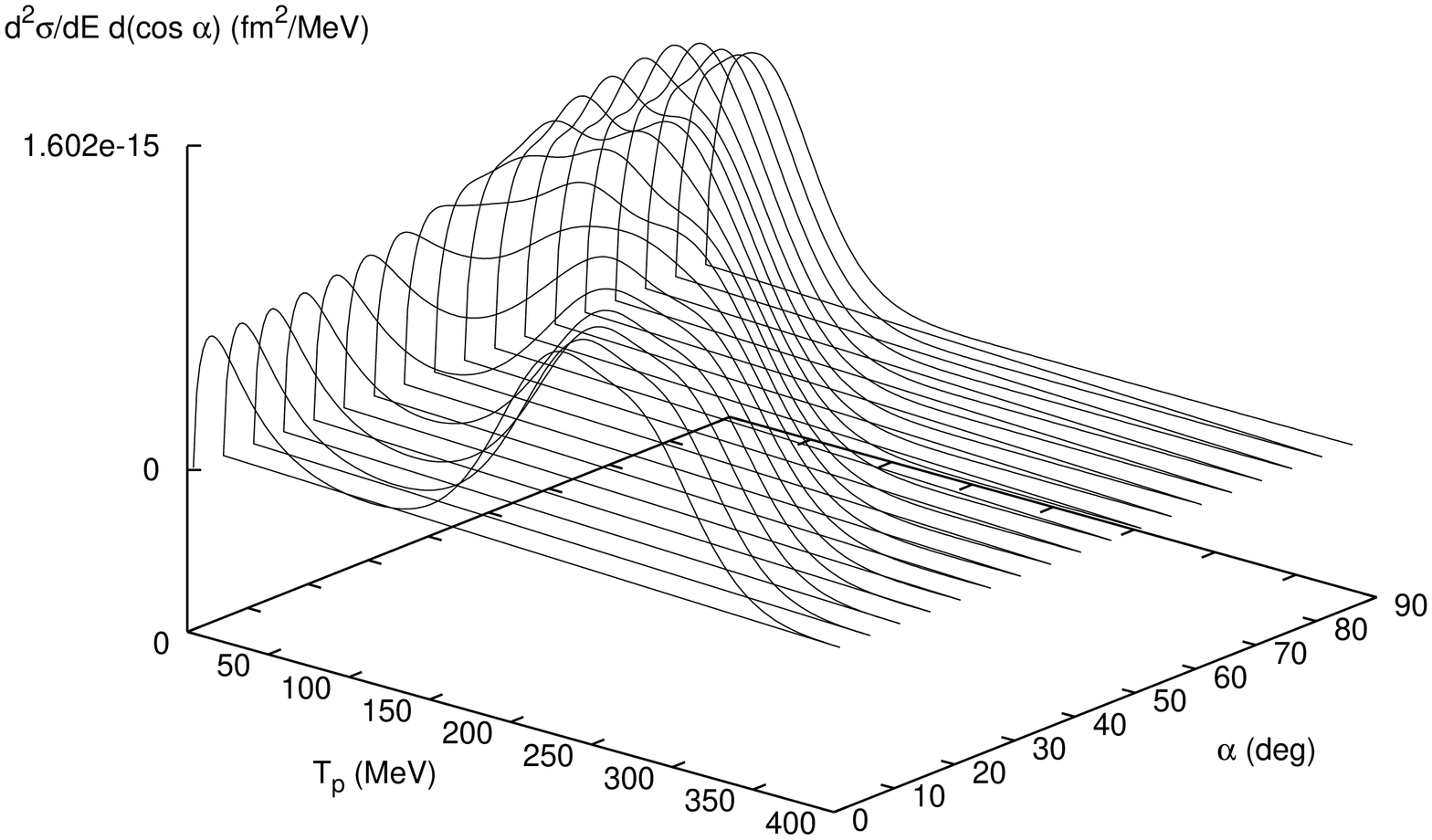}
\caption{Double differential cross section 
 $d^{2} \sigma/dE d (\cos \alpha)$ as a function of the outgoing 
 proton laboratory kinetic energy $T_{p}$ and laboratory scattering 
 angle $\alpha$. The calculation shown is for proton knockout from 
 the $1p^{3/2}$ orbital of $^{12}$C at an incident neutrino energy 
 of 500 MeV.}
 \label{fig_5} 
\end{figure}

\begin{figure}
\includegraphics[height=15cm,angle=-90]{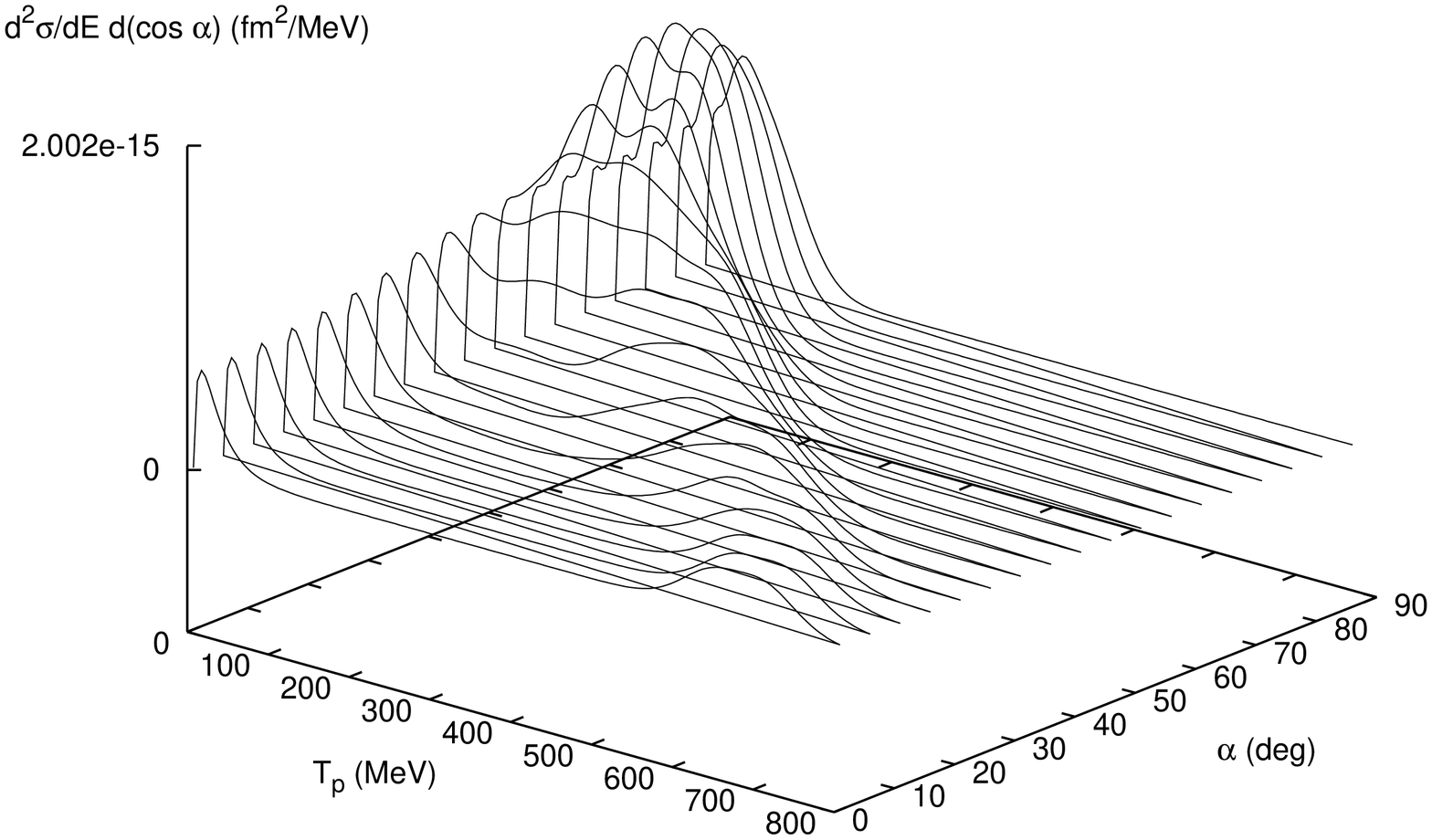}
\caption{Double differential cross section 
 $d^{2} \sigma/dE d (\cos \alpha)$ as a function of the outgoing 
 proton laboratory kinetic energy $T_{p}$ and laboratory scattering 
 angle $\alpha$. The calculation shown is for proton knockout from 
 the $1p^{3/2}$ orbital of $^{12}$C at an incident neutrino energy 
 of 1000 MeV.}
 \label{fig_6}
\end{figure}

In Ref.~\cite{Horowitz_PRC48_93} the strange-quark content of the
nucleon was studied via neutrino-nucleus scattering by using a
relativistic Fermi gas model of the target nucleus. This amounts to
averaging the free neutrino-nucleon cross section over a sharp
momentum distribution for the struck nucleon. In this work we improve
on the above description by employing bound-nucleon wavefunctions
obtained from a relativistic mean-field approximation to the
accurately calibrated NL3 model of Ref.~\cite{Lalazissis_PRC55_97}. 
To quantify the impact of this improvement we display in 
Figs.~\ref{fig_4} to \ref{fig_6} the double differential cross section 
$d^{2}{\sigma}/dE_{p'}d(\cos\alpha)$ as a function of the ejectile 
nucleon kinetic energy $T_{p}$ and its scattering angle $\alpha$ in 
the laboratory frame.  The incident neutrino energy has been fixed in
these plots at 150, 500 and 1000 MeV, respectively. For illustration
purposes---and only for these three graphs---the strange-quark
contribution to the axial-vector form factor ($g_{A}^{s}$) has been
neglected and only proton knockout from the $1p^{3/2}$ orbital of
$^{12}$C is considered.

For the lowest energy neutrinos the cross section displays a single
well developed peak at low $T_{p}$ that monotonically decreases with
increasing scattering angle $\alpha$. For $E_{\nu}\!=\!500$ MeV, our
Fig.~\ref{fig_5} may be compared directly to Fig.~1 of Horowitz and
collaborators~\cite{Horowitz_PRC48_93}. In particular, for a
scattering angle of $\alpha\!=\!20^{\circ}$, our calculation (third
curve along the $\alpha$ direction) also exhibits the characteristic
double-humped structure. For larger values of $\alpha$ the peaks merge
into one and the cross section develops a shape similar to that of
$E_{\nu}\!=\!150$~MeV. Yet, an important difference between the two
sets of calculations is that our cross section does not develop the
sharp features displayed at small angles in
Ref.~\cite{Horowitz_PRC48_93}. This is due to the more realistic
momentum distributions used in our calculations.

\begin{figure}
\includegraphics[height=10cm,angle=0]{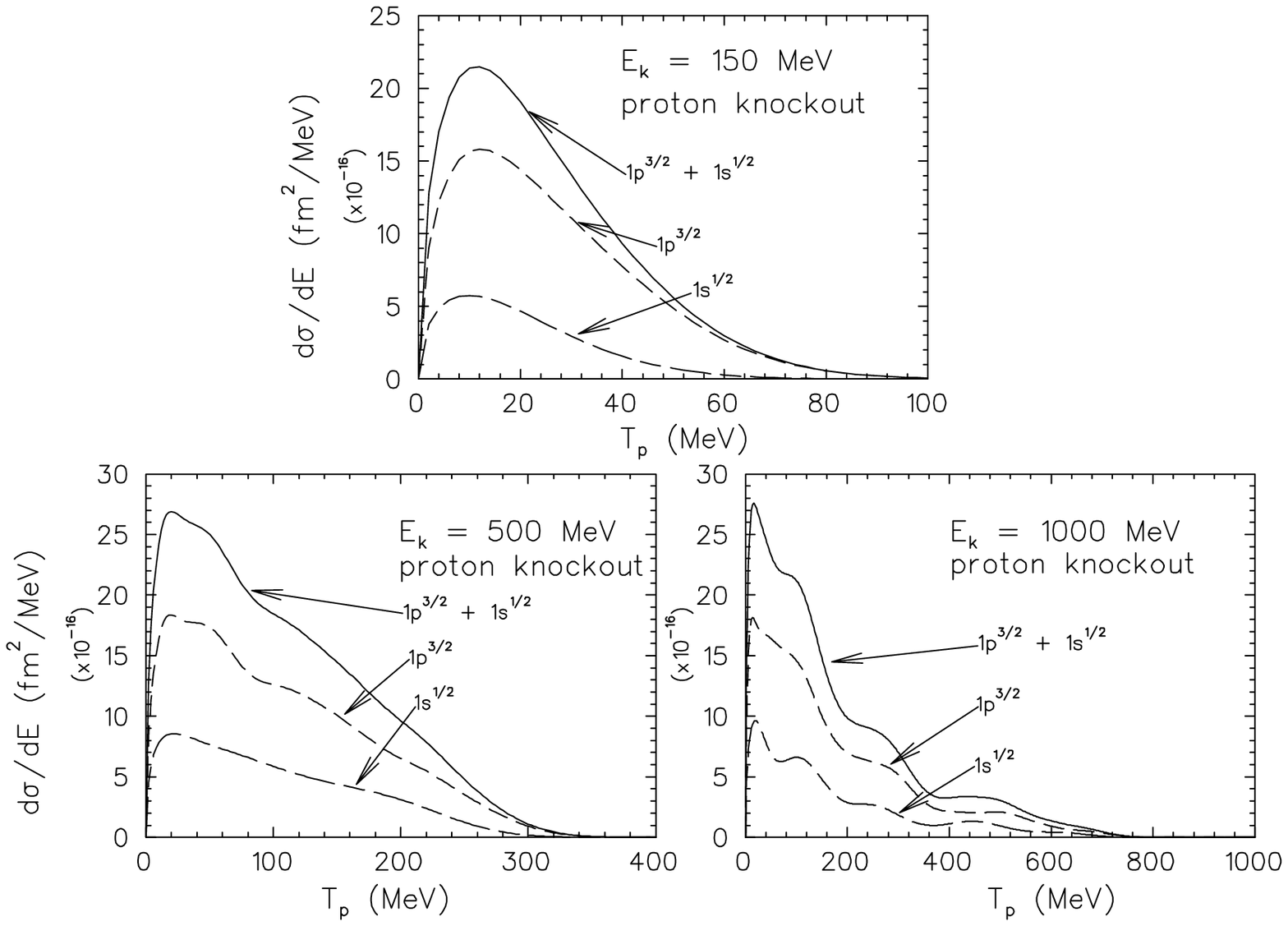}
 \caption{Differential cross section $d \sigma/dE$ as a function 
  of the outgoing proton laboratory kinetic energy $T_{p}$. The 
  dashed and long-dashed--short-dashed lines are for proton 
  knockout from the $1p^{3/2}$ and $1s^{1/2}$ orbitals of $^{12}$C,
  respectively; the solid line represents their sum. The incident
  neutrino energy is taken to be $E_{k}$ = 150, 500 and 1000 MeV.
 }
 \label{fig_7} 
\end{figure}

\begin{figure}
\includegraphics[height=10cm,angle=0]{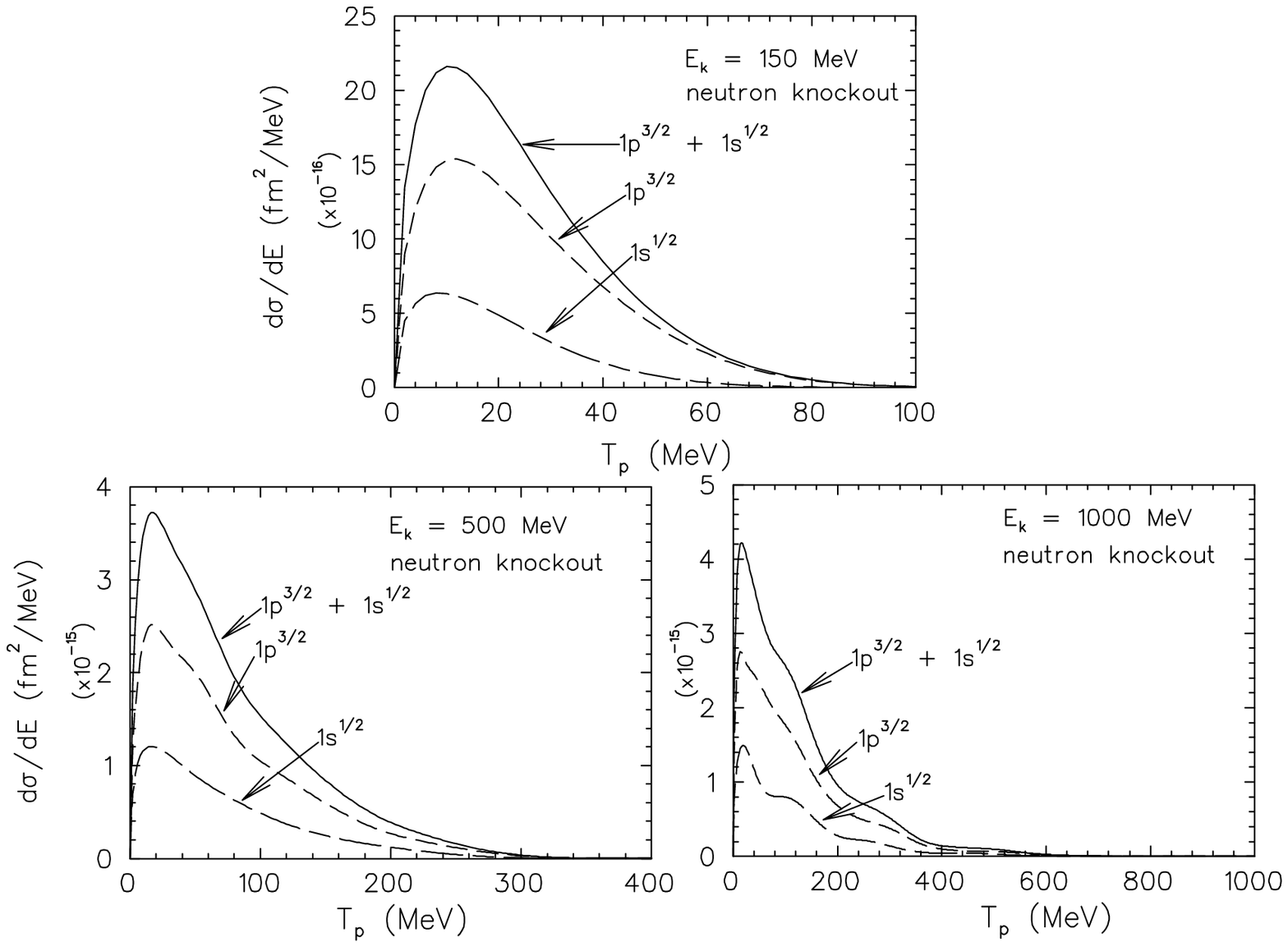}
 \caption{Differential cross section $d \sigma/dE$ as a function 
  of the outgoing neutron laboratory kinetic energy $T_{p}$. The 
  dashed and long-dashed--short-dashed lines are for neutron 
  knockout from the $1p^{3/2}$ and $1s^{1/2}$ orbitals of $^{12}$C,
  respectively; the solid line represents their sum. The incident
  neutrino energy is taken to be $E_{k}$ = 150, 500 and 1000 MeV.
 }
 \label{fig_8} 
\end{figure}

Next, we produce angle-integrated cross sections as a function of
$T_{p}$. Figs.~\ref{fig_7} and \ref{fig_8} display cross sections 
for the knockout of protons and neutrons, respectively. The 
long-dash--short-dashed line represents knockout from the $1s^{1/2}$ 
orbital while the dashed line from the $1p^{3/2}$ orbital of $^{12}$C; 
the solid line displays their sum. As before, calculations are shown 
for incident neutrino energies of 150, 500, and 1000 MeV,
respectively. The cross sections at 150 MeV correspond to those 
shown in Figs. 4 and 5 of Ref.~\cite{Horowitz_PRC48_93}. It has 
been shown in Ref.~\cite{Horowitz_PRC48_93} that binding-energy 
corrections (at 150 MeV) reduce the cross sections relative to
their free Fermi-gas values by about 40\%. Our cross sections 
(already summed over both occupied orbitals) are reduced even 
further. Note that while an average binding energy of 27 MeV has
been used in Ref.~\cite{Horowitz_PRC48_93}, our calculations
include binding energies computed exactly within a mean-field 
approach. At the higher energies of 500 and 1000 MeV the cross 
sections display the same general trend as the one for 150 MeV, 
namely, a peak at a low value of $T_{p}$ and a ``smooth'' falloff 
with increasing $T_{p}$.

\begin{figure}
\includegraphics[height=10cm,angle=0]{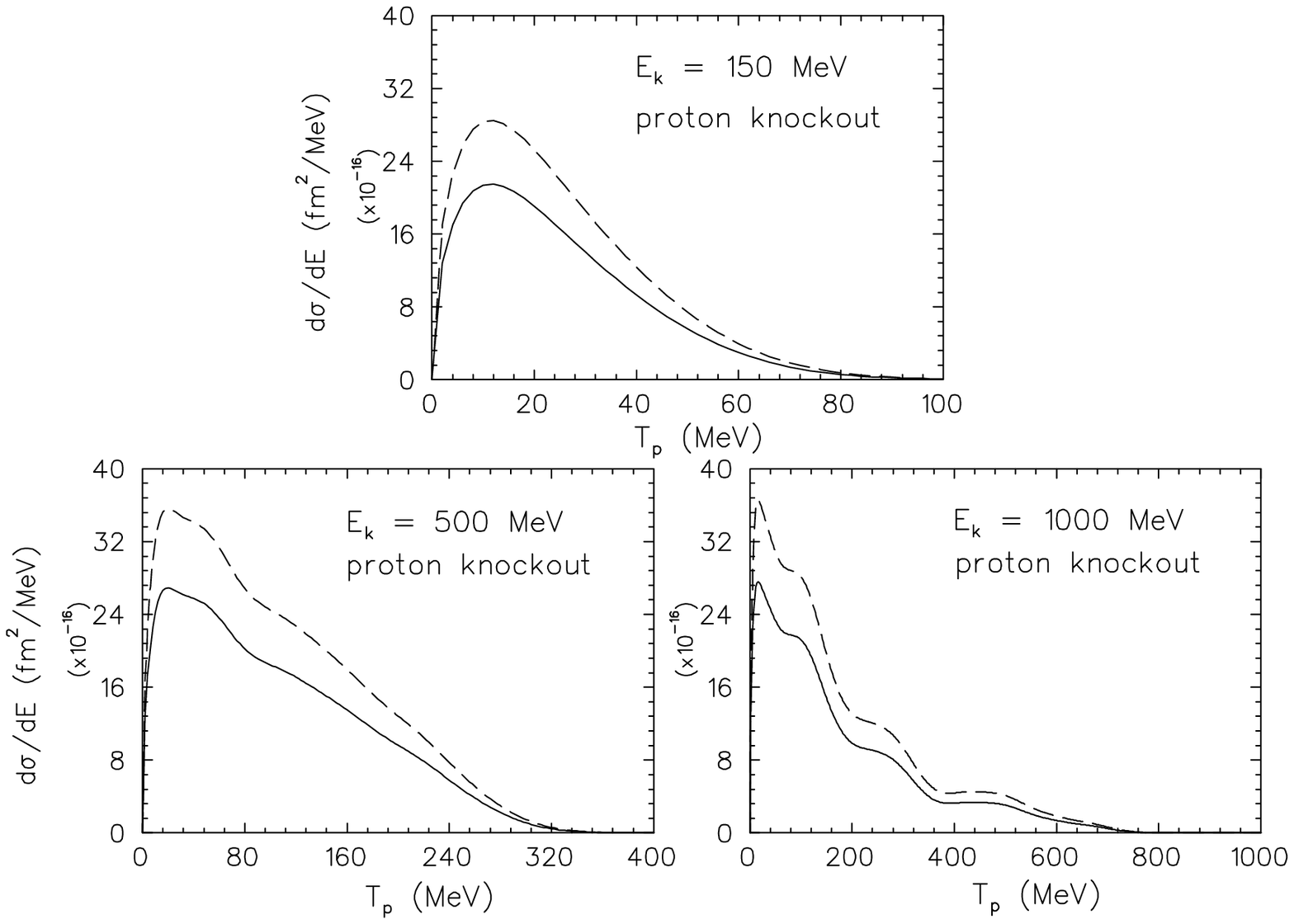}
 \caption{Effect of a strange-quark contribution 
  (of $g_{A}^{s}\!=\!-0.19$) to the axial-vector form factor on 
  the differential cross section $d \sigma/dE$ as a function of 
  the laboratory kinetic energy $T_{p}$ of the outgoing proton. 
  The solid and dashed lines correspond to a zero and a non-zero 
  value of $g_{A}^{s}$, respectively. In this figure we have 
  summed over the $1s^{1/2}$ and $1p^{3/2}$ orbitals of $^{12}$C.
 }
 \label{fig_9} 
\end{figure}

\begin{figure}
\includegraphics[height=10cm,angle=0]{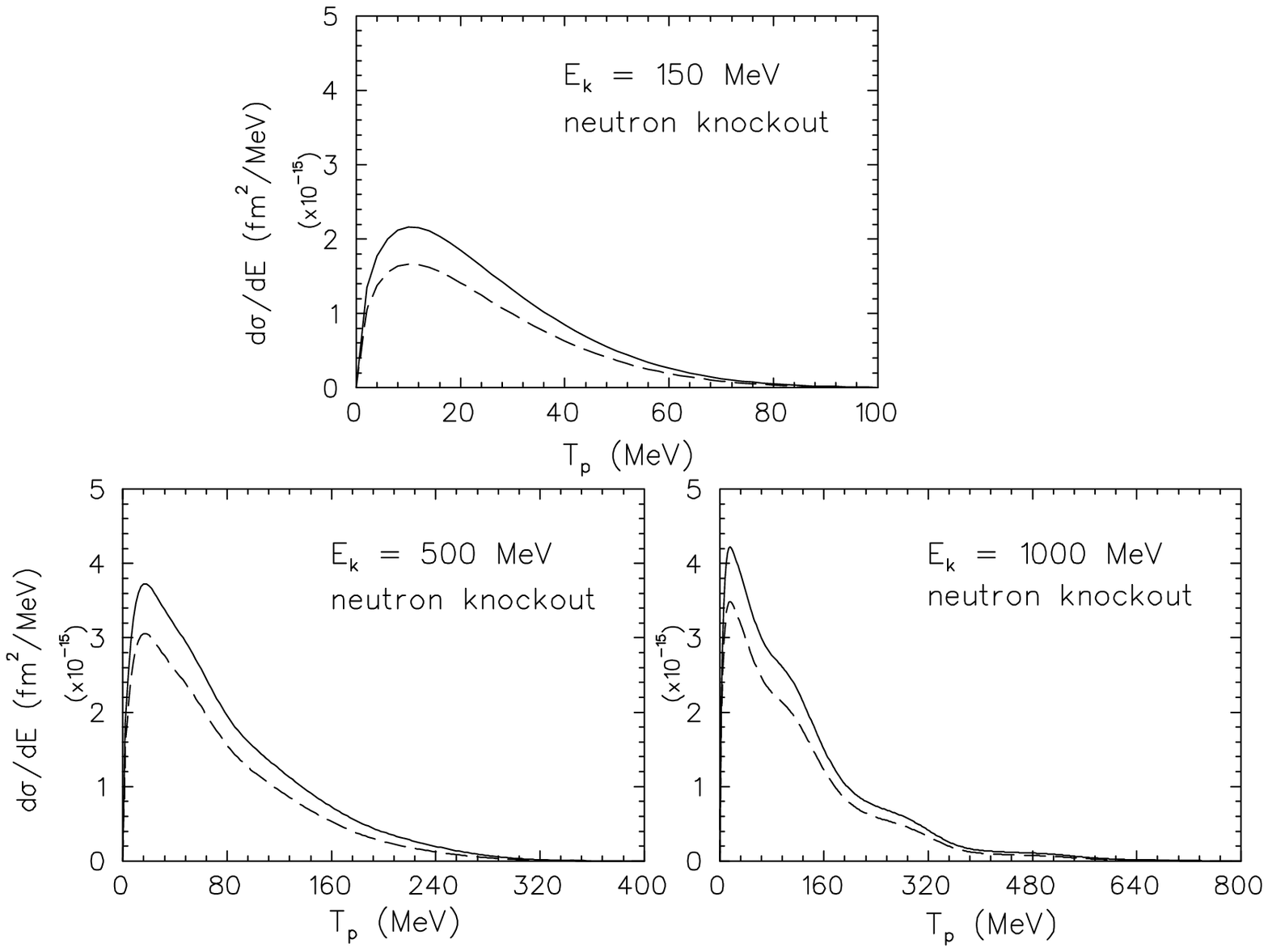}
 \caption{Effect of a strange-quark contribution 
  (of $g_{A}^{s}\!=\!-0.19$) to the axial-vector form factor on 
  the differential cross section $d \sigma/dE$ as a function of 
  the laboratory kinetic energy $T_{p}$ of the outgoing neutron. 
  The solid and dashed lines correspond to a zero and a non-zero 
  value of $g_{A}^{s}$, respectively. In this figure we have 
  summed over the $1s^{1/2}$ and $1p^{3/2}$ orbitals of $^{12}$C.
 }
 \label{fig_10} 
\end{figure}

In Figs.~\ref{fig_9} and \ref{fig_10} we examine the impact of a
strange-quark contribution ($g_{A}^{s}$) to the axial form factor 
on the cross section. Following the discussions in
Refs.~\cite{Ahrens_PRD35_87,Ashman_NPB328_89,Horowitz_PRC48_93} a
value of $g_{A}^{s}\!=\!-\!0.19$ is adopted henceforth. Further, 
in all cases presented here strange-quark contributions to the weak
vector form factors are ignored. The solid and dashed lines correspond
to a zero and a non-zero value of $g_{A}^{s}$, respectively. In both
figures we have summed over the $1s^{1/2}$ and the $1p^{3/2}$ orbitals
of $^{12}$C. Because of the dominance of the axial-vector form factor
(see discussion below) a non-zero value of $g_{A}^{s}$ increases the 
cross section for proton knockout, whereas for neutron knockout the 
cross section is decreased. These findings are consistent with those 
of Refs.~\cite{Horowitz_PRC48_93,Garvey_PLB289_92}.

\begin{figure}
\includegraphics[height=10cm,angle=0]{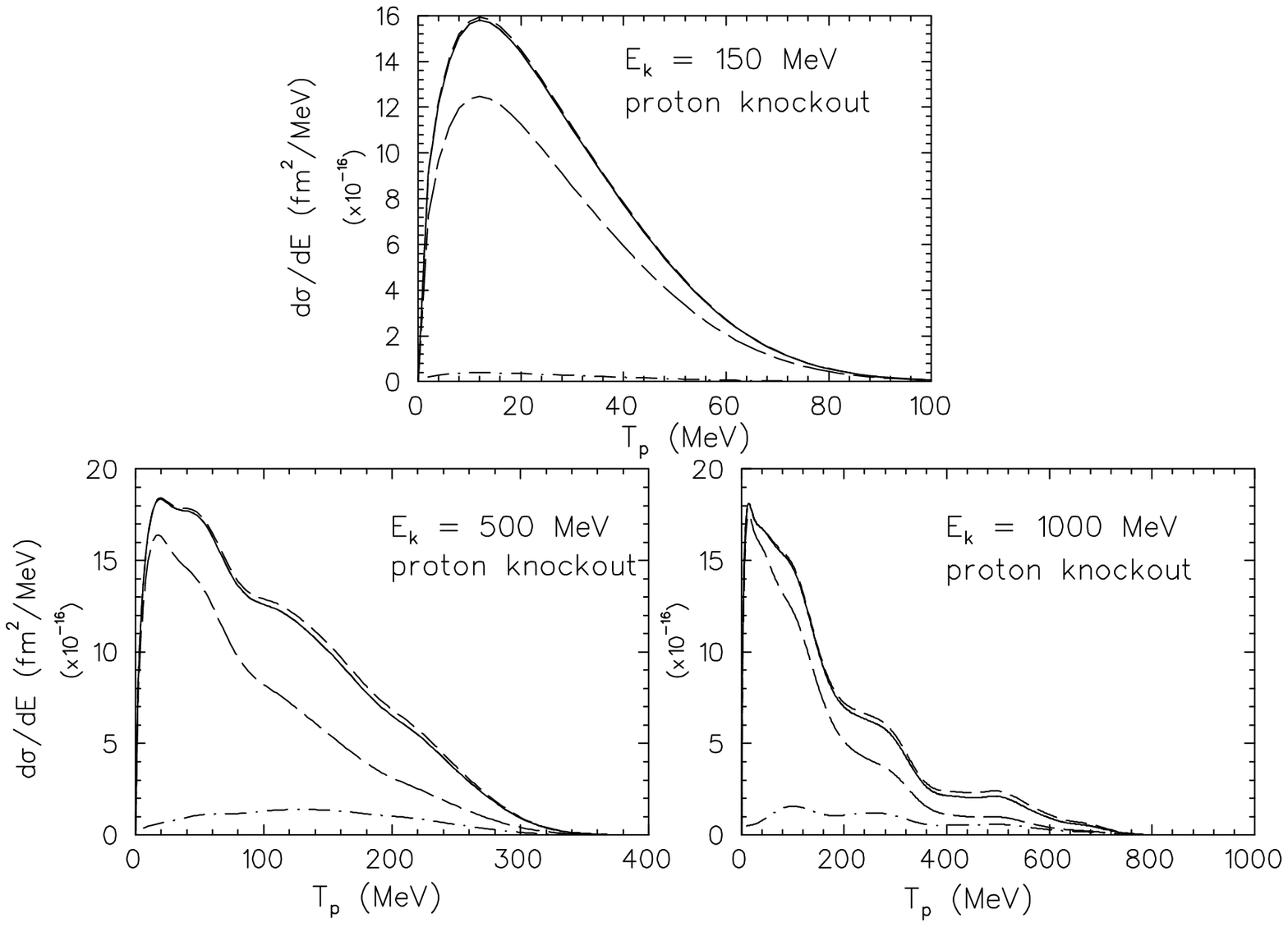}
 \caption{Effect of the single-nucleon form factors on the
  differential cross section $d \sigma/dE$ as a function of the 
  laboratory kinetic energy $T_{p}$ of the outgoing proton. The 
  calculation shown is for proton knockout from only the $1p^{3/2}$ 
  orbital of $^{12}$C at incident neutrino energies of 150, 500 and 
  1000 MeV. Explanation for the various lines is given in the text.}
 \label{fig_11} 
\end{figure}

Next we investigate the role of the various single-nucleon form
factors in the calculation of the differential cross section. 
For this case we restrict ourselves to proton knockout from the 
$1p^{3/2}$ orbital of $^{12}$C. As before, we consider incident 
neutrino energies of 150, 500, and 1000 MeV. The results are shown 
in Fig.~\ref{fig_11}. 
In this figure the solid line corresponds to the case
$(\widetilde{G}_{A}\!\ne\!0, \widetilde{F}_{1}\!\ne\!0$, 
and $\widetilde{F}_{2}\!\ne\!0)$, the dashed line to 
$(\widetilde{G}_{A}\!\ne\!0, \widetilde{F}_{1}\!=\!0$, 
and $\widetilde{F}_{2}\!=\!0)$, the long-dash--short-dashed line to
$(\widetilde{G}_{A}\!\ne\!0, \widetilde{F}_{1}\!=\!0$, and 
$\widetilde{F}_{2}\!\ne\!0)$, and the dash-dot line to 
$(\widetilde{G}_{A}\!=\!0, \widetilde{F}_{1}\!\ne\!0$, and 
$\widetilde{F}_{2}\!\ne\!0)$. For all energies we observe the 
dominant role played by the axial-vector form factor
$\widetilde{G}_{A}$. Indeed, for 150 MeV there is virtually no 
distinction between the calculation using non-zero values for all 
form factors (solid line) and the one where only the axial-vector 
form factor is included (dashed line). This is due to the smallness 
of the weak vector charge of the proton (at all values of $Q^{2}$) and
the low-momentum transfer of the reaction, which makes the
contribution from the Pauli form factor small. The dominance of the
axial-vector form factor is further illustrated by the fact that when
it is set to zero, the cross section becomes vanishingly small
(dash-dot line). This behavior is important as it increases the
sensitivity of the reaction to the strange-quark contribution to the
the axial form factor. Indeed, for 
$\widetilde{F}_{1}\!=\!\widetilde{F}_{2}\!\equiv\!0$, the differential 
cross section becomes proportional to the square of the axial-vector
form factor which, at $Q^{2}\!=\!0$, is given by
(see Appendix~\ref{section_appendix1})
\begin{equation}
\label{eq_82}
  \widetilde{G}_{A}^{2}(Q^{2}\!=\!0)=(g_{A}-g_{A}^{s})^{2}=
  \Big(g_{A}^{2}+(g_{A}^{s})^{2}-2g_{A}g_{A}^{s}\Big) \;.
\end{equation}
The sensitivity to the strange form factor comes about through the 
interference term. 

\begin{figure}
\includegraphics[height=10cm,angle=0]{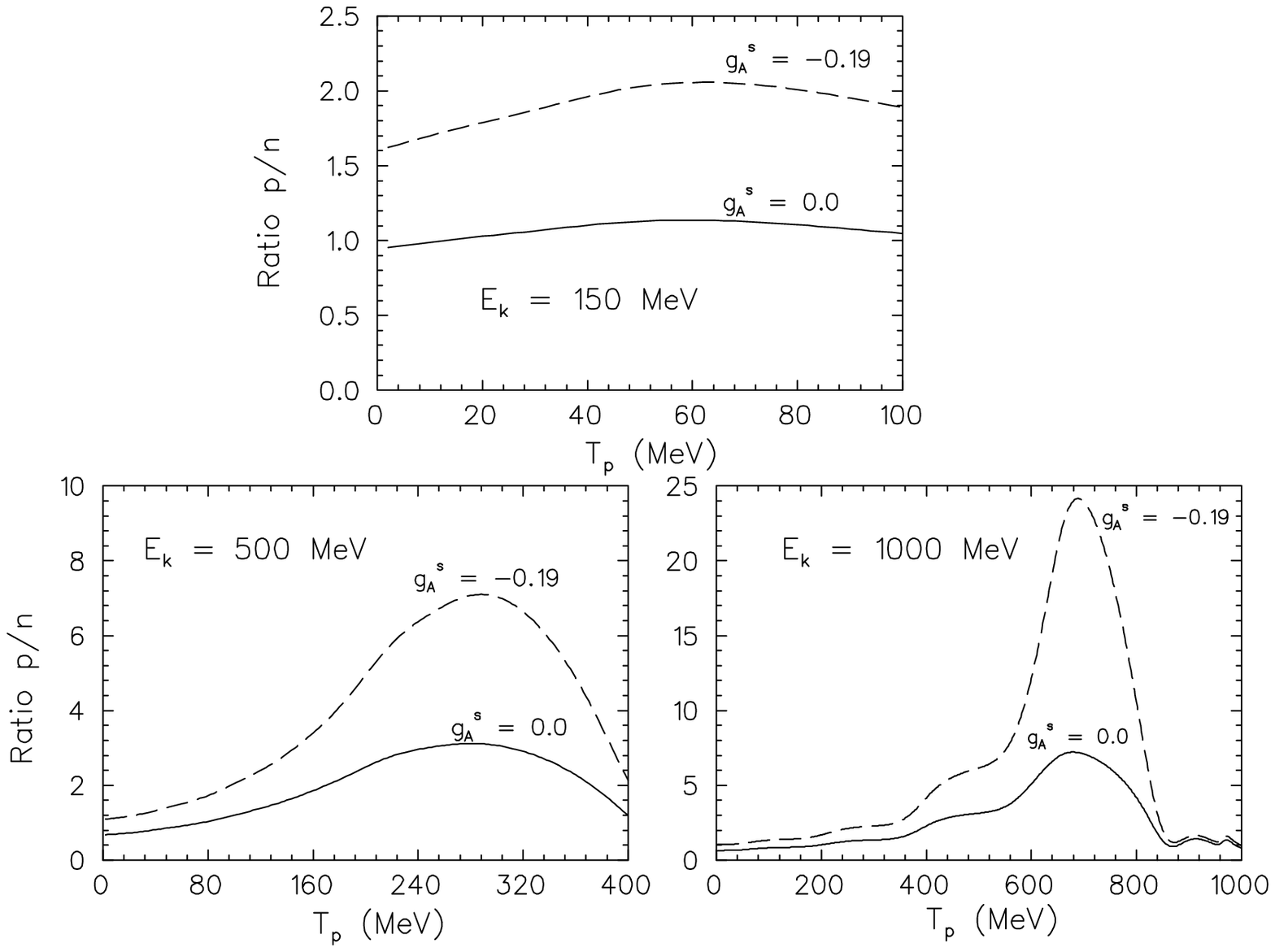}
 \caption{Ratio of cross sections for protons over that for neutrons 
  as a function of the laboratory kinetic energy $T_{p}$ of the 
  outgoing nucleon. Cross sections include contributions from the
  $1s^{1/2}$ and $1p^{3/2}$ orbitals of $^{12}$C.}
 \label{fig_12} 
\end{figure}

An important problem encountered in the previous (and most of the
earlier) analysis is the dependence of the cross section on distortion
effects, that is, on the final-state interactions between the ejectile
nucleon and the residual
nucleus~\cite{Garvey_PLB289_92,Barbaro_PRC54_96}. In an effort to
circumvent this problem, the authors of Ref.~\cite{Garvey_PLB289_92}
proposed to measure the {\it ratio} of quasielastic proton yield
to quasielastic neutron yield, rather than the absolute cross
section. Focusing on the ratio of cross sections proves to be
advantageous for a number of reasons. For example, the calculation of
the angle-integrated cross section in Ref.~\cite{Horowitz_PRC48_93} is
particularly sensitive to binding-energy corrections: both proton and
neutron knockout cross sections are quenched by almost 40\% relative
to the free Fermi-gas estimate. Unfortunately, a strange-quark
contribution to the axial-vector form factor of the neutron also
reduces the cross section by nearly 40\%. Hence, it might be difficult
to separate nuclear-binding effects from a genuine strange-quark
contribution. (We reiterate that, while it remains advantageous to
reduce the sensitivity of the reaction to nuclear-structure effects,
the merit of our calculation is that it incorporates realistic binding 
energies and momentum distributions.) Further, while distortion
effects modify the cross section, they often do so without
a significant redistribution of strength. Thus the ratio of cross
sections, rather than the cross sections themselves, should be less
sensitive to distortion effects. Indeed, in the model of
Ref.~\cite{Garvey_PRC48_93a} it was shown that the ratio of cross
sections was insensitive to distortion effects. Hence, we conclude this
section by displaying in Fig.~\ref{fig_12} the ratio of cross sections
for protons over that for neutrons. While for the lowest value of the
neutrino energy (150 MeV) the ratio remains fairly constant, a
significant dependence on the outgoing nucleon kinetic energy $T_{p}$
is observed for the other two cases. How sensitive this dependence is
to the high-momentum components in the nuclear wavefunction (induced,
for example, by short-range correlations) remains an important open
question.

\section{Summary and conclusions}
\label{section_summary}

Neutral-current neutrino-nucleus cross sections have been computed in
a relativistic plane wave impulse approximation. The aim of this
contribution was to examine the sensitivity of the reaction to the
strange-quark contribution to the axial-vector form factor of the
nucleon. This was done, in part, in response to the Fermilab Intense
Neutrino Scattering Experiment (FINeSE) initiative. In this work we
have followed closely the seminal contributions made to this subject
by various
groups~\cite{Garvey_PLB289_92,Horowitz_PRC48_93,Alberico_NPA623_97}.
Yet, we have improved significantly on them by incorporating nuclear
structure effects through an accurately-calibrated relativistic
mean-field model. Thus, accurate binding energies and nucleon momentum
distributions were employed.

Our results indicate significant quantitative, although minor
qualitative, changes with respect to the relativistic Fermi-gas model
of Horowitz and collaborators~\cite{Horowitz_PRC48_93}. First,
double-differential cross sections displaying sharp features due to a
discontinuous (Fermi gas) momentum distribution get soften by our
choice of mean-field wave functions. However, most of the sharp
features of the Fermi-gas cross section disappear as soon as one
integrates over the scattering angle of the ejected nucleon. At this
point the shape of the cross section becomes largely insensitive to
the choice of momentum distribution. Not so its magnitude. Low-energy
cross sections with binding-energy corrections included in an average
way were shown to be reduced by 40\% relative to the corresponding
free Fermi-gas estimates~\cite{Horowitz_PRC48_93}. The cross sections
reported here, computed with binding energies obtained from a
relativistic mean-field model, yield even smaller cross sections.

In order to reduce the sensitivity of the strange-quark content of the
nucleon to nuclear-structure effects, while at the same time
eliminating the need for an absolute normalization of the cross
section, the authors of Ref.~\cite{Garvey_PLB289_92} have proposed to
use the ratio of proton to neutron yields. Indeed, these authors
demonstrated that the ratio of cross sections is largely insensitive
to distortion effects~\cite{Garvey_PRC48_93a}. Further, binding-energy
corrections, which were responsible for the large (40\%) reduction of
the absolute cross section, become barely visible in the cross-section
ratio~\cite{Horowitz_PRC48_93}. In our case the insensitivity to
final-state interactions entailed an enormous simplification: by
introducing the notion of a bound-nucleon propagator we have exploited
Feynman's trace techniques to develop closed-form, analytic
expressions for the cross section.

In summary, the sensitivity of neutral-current neutrino-nucleus
scattering to the strange-quark content of the axial-vector form
factor of the nucleon was examined. A model-independent formalism
based on a set of eight nuclear structure functions was developed.
On account of both, the notion of a bound-state nucleon
propagator and the insensitivity of the ratio of proton-to-neutron
yields to distortion effects, we computed all nuclear structure
functions in closed form.  Adopting a value for strange-quark 
contribution to the axial-vector form factor of the nucleon of
$g_{A}^{s}\!=\!-0.19$, led to a significant enhancement in the 
proton-to-neutron yields relative to one in which the strange-quark 
contribution was neglected.

\begin{acknowledgments} 
B.I.S.v.d.V gratefully acknowledges the financial support of the
University of Stellenbosch and the National Research Foundation of
South Africa. This material is based upon work supported by the
National Research Foundation under Grant number GUN 2048567
(B.I.S.v.d.V) and by the United States Department of Energy under
Grant number DE-FG05-92ER40750 (J.P.).

\end{acknowledgments} 
\vfill\eject

\appendix
\section{Hadronic weak-neutral current}
\label{section_appendix1}
Neutral current vector and axial-vector matrix elements between
on-shell nucleon states can be parametrized on general grounds 
in terms of four form factors in the following form:
\begin{subequations}
\begin{eqnarray}
 && \langle 
      N({\bf p}',s')|\hat{J}_{\mu}^{\rm NC}(0)|N({\bf p},s)
    \rangle\!=\!\bar{u}({\bf p}',s')
    \left[
    \widetilde{F}_{1}(Q^{2})\gamma_{\mu}+
   i\widetilde{F}_{2}(Q^{2})\sigma_{\mu\nu}\frac{q^{\nu}}{2M}
    \right]{u}({\bf p},s)\;, \\
 && \langle 
      N({\bf p}',s')|\hat{J}_{\mu 5}^{\rm NC}(0)|N({\bf p},s)
    \rangle\!=\!\bar{u}({\bf p}',s')
    \left[
    \widetilde{G}_{A}(Q^{2})\gamma_{\mu}+
    \widetilde{G}_{P}(Q^{2})\frac{q_{\mu}}{M}
    \right]\gamma_{5}{u}({\bf p},s)\;.
\end{eqnarray}
\end{subequations}
The vector current may be decomposed in terms of isoscalar and 
isovector electromagnetic contributions plus an explicit 
strange-quark contribution, which is assumed isoscalar.
Similarly, the axial-vector current may be related to the 
isovector current measured in neutron beta decay plus an
isoscalar strange-quark contribution. That is,
\begin{subequations}
\begin{eqnarray}
  && \hat{J}_{\mu}^{\rm NC}=\big(2-4\sin^{2}\theta_{\rm W}\big)
     \hat{J}_{\mu}^{\rm EM}(T\!=\!1)-4\sin^{2}\theta_{\rm W}
     \hat{J}_{\mu}^{\rm EM}(T\!=\!0)-\bar{s}\gamma_{\mu}s \;, \\
  && \hat{J}_{\mu 5}^{\rm NC}=-2\hat{A}_{\mu}(T\!=\!1)
                       +\bar{s}\gamma_{\mu}\gamma_{5}s \;. 
\end{eqnarray}
\end{subequations}
This decomposition enables one to express the two vector and
the one axial-vector form factors in the following form: 
\begin{subequations}
\begin{eqnarray}
  && \widetilde{F}_{i}(Q^{2})=
     \big(1-4\sin^{2}\theta_{\rm W}\big)F_{i}^{(p)}(Q^{2})-
      F_{i}^{(n)}(Q^{2})-F_{i}^{(s)}(Q^{2})\;; \quad 
      {\rm for \, protons} \;, \\
  && \widetilde{F}_{i}(Q^{2})=
     \big(1-4\sin^{2}\theta_{\rm W}\big)F_{i}^{(n)}(Q^{2})-
      F_{i}^{(p)}(Q^{2})-F_{i}^{(s)}(Q^{2})\;; \quad 
      {\rm for \, neutrons} \;, 
\end{eqnarray}
\end{subequations}
and
\begin{subequations}
\begin{eqnarray}
 && \widetilde{G}_{A}(Q^{2})=+{G}_{A}^{(3)}(Q^{2})-
    {G}_{A}^{(s)}(Q^{2})\;; \quad {\rm for \, protons} \;, \\
 && \widetilde{G}_{A}(Q^{2})=-{G}_{A}^{(3)}(Q^{2})-
    {G}_{A}^{(s)}(Q^{2})\;; \quad {\rm for \, neutrons} \;.
\end{eqnarray}
\end{subequations}
Note that because (massless) neutrino scattering is insensitive
to the induced pseudoscalar form factor $\widetilde{G}_{P}$, it
has been ignored throughout this work. In what follows, standard 
parameterizations of the electromagnetic and axial-vector nucleon 
form factors are employed. In particular, we follow the conventions 
adopted in Ref.~\cite{Musolf_PhysRep239_94}. These are given by
\begin{subequations}
\begin{eqnarray}
 && F^{(p)}_{1}(Q^{2})=
    \left(\frac{1+\tau(1+\lambda_{p})}{1+\tau}\right)
    G_{D}^{V}(Q^{2})\;, \quad
    F^{(p)}_{2}(Q^{2})= 
    \left(\frac{\lambda_{p}}{1+\tau}\right)G_{D}^{V}(Q^{2})\;, \\
 && F^{(n)}_{1}(Q^{2})=
    \left(\frac{\lambda_{n}\tau(1-\eta)}{1+\tau}\right)
    G_{D}^{V}(Q^{2})\;, \quad
    F^{(n)}_{2}(Q^{2})=
    \left(\frac{\lambda_{n}(1+\tau\eta)}{1+\tau}\right)
    G_{D}^{V}(Q^{2})\;, \\
 && G^{(3)}_{A}(Q^{2})=g_{A}G_{D}^{A}(Q^{2})\;, \quad
    G^{(s)}_{A}(Q^{2})=g^{s}_{A}G_{D}^{A}(Q^{2})\;,
\end{eqnarray}
\end{subequations}
where a dipole form factor of the following form is assumed:
\begin{subequations}
\begin{eqnarray}
  && G_{D}^{V}(Q^{2})=(1+Q^{2}/M_{V}^{2})^{-2}=(1+4.97\tau)^{-2}\; \\
  && G_{D}^{A}(Q^{2})=(1+Q^{2}/M_{A}^{2})^{-2}=(1+3.31\tau)^{-2}\; \\
  && \eta=(1+5.6\ \tau)^{-1} \quad \tau=Q^{2}/(4M^{2})\;.
\end{eqnarray}
\end{subequations}
Finally, for reference we display the value of the various nucleon
form factors at $Q^{2}\!=\!0$
\begin{subequations}
\begin{eqnarray}
 && F_{1}^{(p)}(0)=1\;, \quad
    F_{1}^{(n)}(0)=0\;, \quad
    F_{1}^{(s)}(0)=0\;, \\
 && F_{2}^{(p)}(0)=\lambda_{p}=+1.79\;, \quad
    F_{2}^{(n)}(0)=\lambda_{n}=-1.91\;, \quad
    F_{2}^{(s)}(0)=\mu_{s}\;, \\
 && G_{A}^{(3)}(0)=g_{A}=+1.26\;, \quad
    G_{A}^{(s)}(0)=g_{A}^{s}\;. 
\end{eqnarray}
\end{subequations}
\section{Hadronic Tensor in a RPWIA}
\label{section_appendix2}

The hadronic tensor computed in a relativistic plane-wave
impulse approximation has been written in Eq.~(\ref{eq_62})
in terms of eight structure functions. These are given by
the following simple forms:
\begin{subequations}
\begin{eqnarray}
 \nonumber
 \widetilde{W}_{1} & = &
 4\widetilde{F}_{1}^{2}(MM_{\alpha}-p_{\alpha}\cdot p')+
 \frac{\widetilde{F}_{2}^{2}}{M^{2}} 
 \Big[(MM_{\alpha}+p_{\alpha}\cdot p')q^{2}-
 2(p_{\alpha}\cdot q)(p'\cdot q)\Big]  \\
 \label{eq_63}
 & - &
 4\widetilde{G}_{A}^{2}(MM_{\alpha}+p_{\alpha}\cdot p')+
 \frac{4\widetilde{F}_{1}\widetilde{F}_{2}}{M} 
 \Big[(p'\cdot q)M_{\alpha}-(p_{\alpha}\cdot q)M\Big] \;, \\
 \label{eq_64}
  \widetilde{W}_{2} & = &
 4\Big(\widetilde{F}_{1}^{2}+\widetilde{G}_{A}^{2}\Big)- 
  \widetilde{F}_{2}^{2}\left(\frac{q^{2}}{M^{2}}\right) \;, \\
 \label{eq_65}
  \widetilde{W}_{3} & = &
  \widetilde{F}_{2}^{2}\left(\frac{p'\cdot q}{M^{2}}\right)+
 2\widetilde{F}_{1}\widetilde{F}_{2} \;, \\
 \label{eq_66}
  \widetilde{W}_{4} & = &
  \widetilde{F}_{2}^{2}\left(\frac{p_{\alpha}\cdot q}{M^{2}}\right)-
 2\widetilde{F}_{1}\widetilde{F}_{2}\left(\frac{M_{\alpha}}{M}\right)
  \;, \\
\label{eq_67}
 \widetilde{W}_{5} & = & -
 \frac{\widetilde{F}_{2}^{2}}{M^{2}}
 (MM_{\alpha}+p_{\alpha}\cdot p') \;, \\
\label{eq_68}
 \widetilde{W}_{6} & = & 8i\widetilde{F}_{1}\widetilde{G}_{A} \;, \\
\label{eq_69}
 \widetilde{W}_{7} & = & 4i\widetilde{F}_{2}\widetilde{G}_{A} \;, \\
\label{eq_70}
 \widetilde{W}_{8} & = & 4i\left(\frac{M_{\alpha}}{M}\right)
 \widetilde{F}_{2}\widetilde{G}_{A} \;.
\end{eqnarray}
\end{subequations}

Our final task consists in relating the structure functions in the 
general model-independent expansion of $W^{\mu\nu}$ given in 
Eqs.~(\ref{eq_29}) and~(\ref{eq_30}), to the model-dependent ones 
given above. This is done by noting that in the laboratory frame 
one can write
\begin{equation}
\label{eq_71}
 p_{\alpha}^{\mu} = a P^{\mu} + b (p'^{\mu} - q^{\mu}) \;,
\end{equation}
where the $a$ and $b$ coefficients are defined as follows:
\begin{equation}
\label{eq_72}
  a \equiv \frac{1}{M_{A}}
  \left(
   E_{\alpha}-\frac{|{\bf p}_{\alpha}|}{|{\bf p}_{m}|}E_{m}
  \right)\;, \quad
  b \equiv \frac{|{\bf p}_{\alpha}|}{|{\bf p}_{m}|} \;.
\end{equation}
Substitution of Eq.~(\ref{eq_72}) into Eq.~(\ref{eq_62}) allows us 
to identify the contribution of $\widetilde{W}_{i}$ to each of the 
structure functions. These are given by
\begin{subequations}
\begin{eqnarray}
\label{eq_73}
 W_{1} & = & \left(\frac{2j+1}{2E_{{\bf p}^{\prime}}}\right)
 \widetilde{W}_{1} \;, \\
 W_{4} & = & \left(\frac{2j+1}{2E_{{\bf p}^{\prime}}}\right)
 \left(2b\widetilde{W}_{2}\right) \;, \\
 W_{7} & = & \left(\frac{2j+1}{2E_{{\bf p}^{\prime}}}\right)
 \left(a\widetilde{W}_{2}\right) \;, \\
 W_{11} & = & \left(\frac{2j+1}{2E_{{\bf p}^{\prime}}}\right)
 \left(-a\widetilde{W}_{7}\right) \;, \\
 W_{12} & = & \left(\frac{2j+1}{2E_{{\bf p}^{\prime}}}\right)
 \left(-b\widetilde{W}_{6}-b\widetilde{W}_{7}-
         \widetilde{W}_{8}\right) \;, \\
 W_{13} & = & \left(\frac{2j+1}{2E_{{\bf p}^{\prime}}}\right)
 \left(-a\widetilde{W}_{6}\right) \;.
\end{eqnarray}
\end{subequations}
Note that the model predicts $W_{3}\!=\!W_{10}\!\equiv\!0$.
Moreover, as was mentioned previously, (massless) neutrino 
scattering is insensitive to the following five structure
functions: $W_{2}$, $W_{5}$, $W_{6}$, $W_{8}$, and $W_{9}$. 
\vfill\eject

\bibliography{NuP} 
\end{document}